\documentclass[nofootinbib,nobibnotes,amsmath,amssymb,amsfonts,noshowpacs,floatfix,notitlepage,onecolumn]{revtex4-1}
\usepackage{cleveref}
\usepackage{graphicx}
\usepackage{amssymb}
\usepackage{xcolor}
\usepackage{booktabs}
\usepackage{multirow}
\usepackage{enumitem,kantlipsum}
\usepackage[multiple]{footmisc}
\usepackage{booktabs}

\usepackage{array,multirow}

\usepackage{amsmath}
\usepackage{mathtools}
\usepackage{cleveref}

\begin{document}
\title{Filament flexibility enhances power transduction of F-actin bundles.}
\author{Alessia Perilli}
\email{alessia.perilli@roma1.infn.it}
\affiliation{Department of Physics, Sapienza University of Rome, P.le Aldo Moro 2, I-00185 Rome, Italy\\ Department of Chemistry, \'Ecole Normale Superi\'eure, rue Lhomond 24, 75005 Paris, France}
\author{Carlo Pierleoni}
\email{carlo.pierleoni@aquila.infn.it}
\affiliation{DSFC, University of L'Aquila, 67100  L'Aquila, Italy\\
Maison de la Simulation, CEA, CNRS, Universit\'e Paris-Sud, UVSQ, Universit\'e Paris-Saclay, 91191 Gif-sur-Yvette, France.}
\author{Jean-Paul Ryckaert}
\email{jryckaer@ulb.ac.be}
\affiliation{Physics Dept., Universit\'e Libre de Brussels (ULB), Campus Plaine, CP 223, 
B-1050 Brussels, Belgium}

\date{\today}

%
%
%

\begin{abstract}
The dynamic behavior of bundles of actin filaments growing against a loaded obstacle is investigated through a generalized version of the standard multi filaments Brownian Ratchet model in which the (de)polymerizing filaments are treated not as rigid rods but as semi-flexible discrete wormlike chains with a realistic value of the persistence length.
 By stochastic dynamic simulations we study the relaxation of a bundle of $N_f$ filaments with staggered seed arrangement against a harmonic trap load in supercritical conditions. Thanks to the time scale separation between the wall motion and the filament size relaxation, mimiking realistic conditions, this set-up allows us to extract a full load-velocity curve from a single experiment over the trap force/size range explored. We observe a systematic evolution of steady non-equilibrium states over three regimes of bundle lengths $L$. A first threshold length $\Lambda$ marks the transition between the rigid dynamic regime ($L<\Lambda$), characterized by the usual rigid filament load-velocity relationship $V(F)$, and the flexible dynamic regime ($L>\Lambda$), where the velocity $V(F,L)$ is an increasing function of the bundle length $L$ at fixed load $F$, the enhancement being the result of an improved level of work sharing among the filaments induced by flexibility.  A second critical length corresponds to the beginning of an unstable regime characterized by a high probability to develop escaping filaments which start growing laterally and thus do not participate anymore to the generation of the polymerization force. This phenomenon prevents the bundle from reaching at this critical length the limit behavior corresponding to Perfect Load Sharing.
\end{abstract}
\maketitle

\section{Introduction}
\label{sec:Intro}
Cellular shape changes involve significant correlated deformations of the cell membrane overlying and embedding the cytoskeleton. In the particular case of lamellipodium and filopodium growth the nanoscale energy source for such deformation is provided by actin polymerization: G-actin monomers assemble to the barbed end of filaments in direct contact with the cell membrane, generating significant pushing forces. 
The speed of membrane deformation/displacement depends on the resisting load due to membrane deformation and to the crowded environment around the cell \cite{Falcke.12}. 
In the specific case of filopodium growth, modelling has been developed over the last twenty years in different directions. 
Mogilner and Rubinstein \cite{Mogilner.Rubinstein} pioneered the direct modeling of a filopodium protruding against the membrane resistance, establishing conditions for its onset and general size limitations due to actin bundle buckling or to the diffusion of G-actin monomers to the tip. 
A full stochastic model of filopodium growth based on a bundle of a few rigid filaments taking into account retrograde flow, a redistribution of the total load on filament tips induced by membrane fluctuations and a rather realistic treatment of monomer diffusion which couples to polymerization events was used to unravel the different length and time scales associated with filopodium growth.  In particular, the growth rate of filaments and their stable sizes were related to various parameters, like the G-actin diffusivity or the amplitude of membrane fluctuations\cite{Papoian.08}.
In a subsequent refined model with explicit G actin particles interacting with the rigid filaments via excluded volume interactions, it was found that the restricted diffusion of G-actin taking place in the enclosing cylindric volume of the membrane tube with growing filaments as local obstacles, leads to an inhomogeneous growth of the bundle which favors elongation of peripheral filaments with respect to filaments located near the centre of the tube \cite{Papoian.16}. 
Further, alternative ways to model membrane fluctuations and probe their effect on the bundle growth velocity was specifically addressed by including in stochastic models membrane height fluctuations at the filament tips \cite{AWS06,Sadhu.16}.

\begin{figure}[ht]
\begin{center}
\includegraphics[width=0.6\columnwidth]{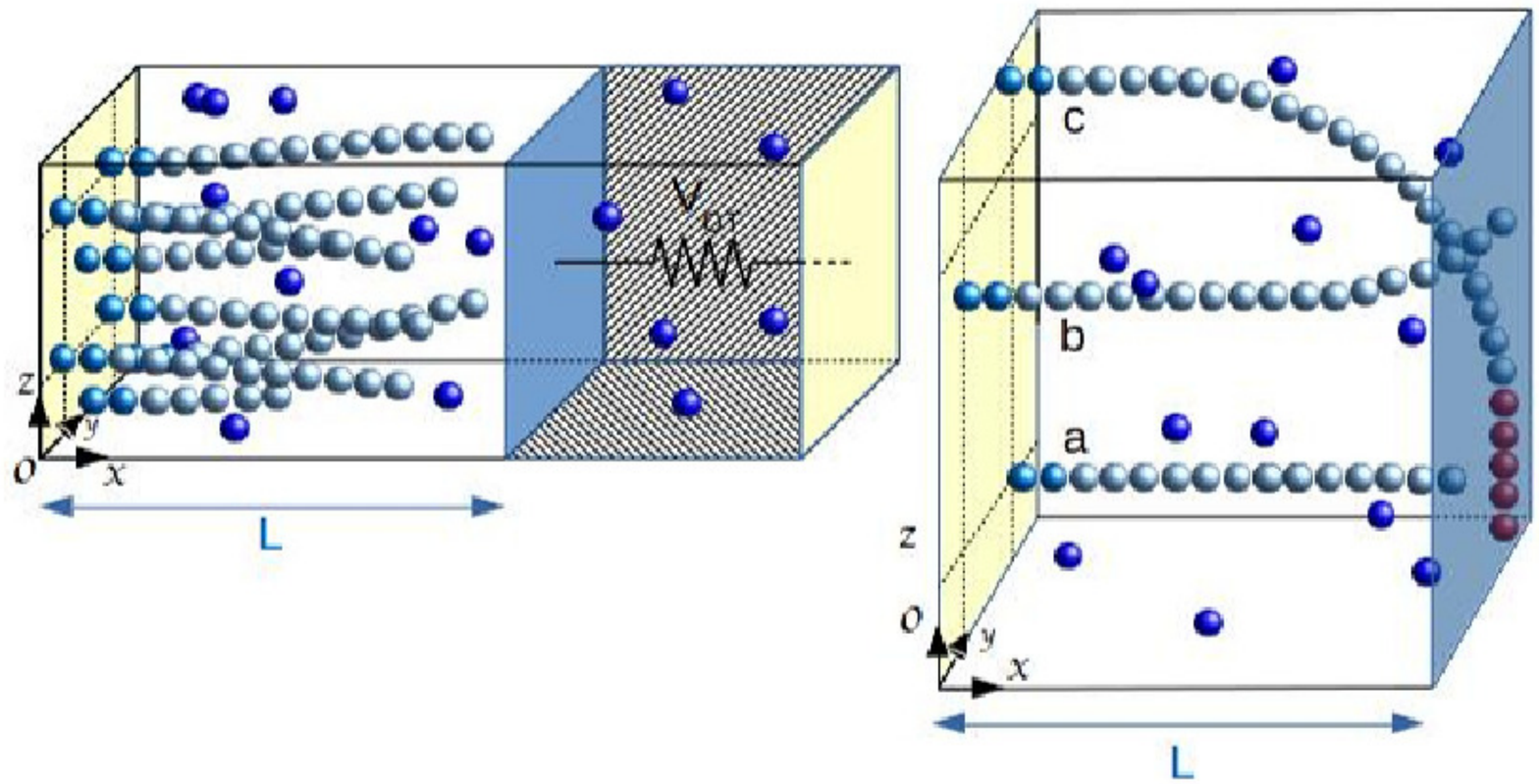}
\caption{Left panel: Bundle of mutually non-interacting flexible filaments grafted normally to the fixed wall on the left, pressing on a mobile wall (blue plane) oriented perpendicular to the grafting direction. The latter wall, located at position $x=L$ and subject to an harmonic load $F=-\kappa_T L$ (mechanical potential energy $V_{OT}=\frac{1}{2}k_T L^2$), separates two chambers both in contact with a reservoir of monomers (dark blue spheres) at fixed concentration. Light blue spheres depict monomers incorporated into grafted F-actin filaments in the left chamber while the second chamber contains free monomers only. This illustrates the underlying ``microscopic" model, treated recently by equilibrium statistical mechanics in the grand canonical ensemble \cite{PPCR.16}, from which our coarse-grained dynamical stochastic model derives. 
\\Right panel: Illustrations of (a) a grafted filament with size $z(L)=1+\text{int}(L/d)$ (where int($x$) represents the integer part of $x$) corresponding to the longer size for which no interactions  with the wall at $L$ are possible; (b) a slightly bent grafted filament with net polymerization arrested by the wall; (c) an escaping filament polymerizing along the obstacle while subject to a bending mechanical force by the wall. A filament is considered as escaping when its size in monomers exceeds the threshold $z^*(L)={\text{int}}(\frac{\pi L}{2d}))$ for which the filament forms a quarter of a circle of radius equal to the trap width $L$. In the figure, the monomers in excess to $z^*$ are shown in red.} 
\label{fig:sys}
\end{center}
\end{figure}

Going one step further in dissecting different effects, focus is often brought on the basic mechanisms and the collective strategies put in place by bundles of actin filaments to optimize the transduction of chemical into mechanical energy\cite{Hill.81}. Grafted actin filaments in a solution of free G-actin monomers at density $\rho_1$ polymerize, with a rate $U_0=k_{on} \rho_1$, and depolymerize, with a rate $W_0=k_{off}$, at their free ends, where $k_{on}$ and $k_{off}$ are kinetic constants. At the bulk critical density $\rho_{1c}=k_{off}/k_{on}$ the filaments neither grow nor shrink on average. Supercritical conditions $\rho_1>\rho_{1c}$ correspond to a net filament growth while subcritical conditions ($\rho_1<\rho_{1c}$) to a net filament shrinkage. The velocity of elongation of the filaments in absence of obstacles is $V_0=d(U_0-W_0)$ where $d=2.7 nm$ is the filament length increment per added monomer. When growing against a loaded wall in supercritical conditions, the net velocity $V<V_0$ of the wall times the external load $F$ is a measure of the power transduction of chemical energy into mechanical work provided by growing pushing filaments. When filaments push on the wall at normal incidence, the wall velocity decreases for increasing loads from $V_0$ at zero load \footnote{Taking into account viscous forces on the wall motion leads to a value slightly smaller than $V_0$\cite{BM.05}.}. For strong enough loads the filaments will be forced to depolymerize on average and the net wall velocity will become negative even in supercritical conditions. The stalling force $F_s$ is the specific value of the load which stops the obstacle ($V=0$) and hence equals the bundle polymerization force at rest.

On the experimental side, measurements of the polymerization force exerted by a bundle of actin filaments on an obstacle and the corresponding load-velocity relation are difficult to extract in \textit{in-vivo} conditions given the uncertainties on various possible additional effects e.g. those linked to interferences of auxiliary proteins (see however ref. \cite{BRVJvNB2013}). \textit{In-vitro} experiments targeting dynamic bundle properties exploit actin bundles glued on colloidal particles growing against a load, the obstacle being another colloidal particle or a fixed wall \cite{Footer.07,DCBB.14}. The theoretical framework used to interpret such experiments is usually provided by some form of multi filament Brownian Ratchet (BR) models of fully \textit{rigid} filaments polymerizing against a loaded rigid wall \cite{POO.93,Sander.20,Joanny.11,Mogilner.99,Kierfeld.2011,Kolomeisky.2015}. The rigid filament hypothesis is assumed to be valid because typical F-actin (uncrosslinked) filaments rarely reach contour lengths $L_c$ longer than a few $0.1\mu m$ which is much smaller that the persistence length $\ell_p \approx 15 \mu m$. To be able to grow normally against a loaded wall while pushing, straight rigid filaments need to exploit temporary gaps of size $\geq d$ created by thermal fluctuations of the wall, allowing monomer insertion and thus the targeted rectification of the biased random motion of the obstacle.

The rigid character of the filaments makes these models unidimensional while in reality F-actin filaments are semi-flexible and this leads to two distinct contour length dependent features for their compressional behavior. First, a new mechanism of generation of voids of size $d$ as a result of tip bending fluctuations becomes possible, as proposed by the Elastic Brownian Rachet (EBR) model \cite{Mogilner.96}. Using this EBR model, only a moderate velocity increase in the load-velocity relation of a single filament pushing a hard wall has been evidenced by numerical simulations \cite{BM.05,BM.06}. Next, in absence of the lateral envelop around a bundle of uncrosslinked polymerizing filaments (as provided by the membrane tube in \textit{in-vivo} filopodia \cite{Blanchoin.14} or in similar finger-like structures reconstructed \textit{in-vitro} \cite{Fletcher.08}), filaments in a long enough bundle may start growing in lateral directions rather than pushing the obstacle. Such diverging filaments are said to be escaping as they do not participate anymore to the direct transformation of chemical energy into mechanical work against the load, even if a mechanical force is still exerted by the wall to maintain the permanent filament bending. The escaping filament phenomenon, also called the pushing catastrophe \cite{BM.05,BM.06}, is minimal, but nevertheless present, when semi-flexible filaments hit the wall at normal incidence, that is in the geometry illustrated in figure \ref{fig:sys}. 
Simulations at constant load with the EBR model\cite{BM.05} found for escaping filaments a probability of occurrence per unit of time increasing as the square of the contour length of the grafted filament. 
The presence of escaping filaments at normal incidence was detected in \textit{in-vitro} experiments when following actin bundle growth against an optical trap load \cite{Footer.07} within the geometry illustrated in figure \ref{fig:sys}\footnote{In this reference the escaping phenomenon is confused with macroscopic buckling.}. Measurements of the load-velocity of actin bundles obtained by monitoring elongation velocity against a constant load \cite{DCBB.14} have been interpreted with the multi-filament BR model with staggered filament seeds disposition: whether the flexibility features of filaments are negligible because the filaments remain sufficiently short or whether these effects renormalize parameters (like the number of filaments in the bundle) in the fitting procedure is difficult to be established. 

As we develop below and exploit in the present paper, a refined theoretical analysis of the filament flexibility effects is facilitated by considering the actin bundle in an optical trap, following the setup exploited in \cite{Footer.07}. Figure \ref{fig:sys} represents the situation of a thought experiment of a bundle of polymerizing semi-flexible filaments in an optical trap apparatus, there represented by a harmonic resisting load (see the caption of the figure). The moving wall (blue in the figure) will have its rest position calibrated at the left yellow wall ($L=0$), in absence of filaments. The left yellow wall is also the location where filaments seeds are grafted. In presence of free monomers in the solution in supercritical conditions, the growing filaments will push the blue wall in the right direction but the optical trap will increase its reaction linearly with the position of the blue wall ($F=-\kappa_T L$) until stalling is reached, which is the condition at which the velocity of the wall vanishes. The intensity of the trap $\kappa_T$ tunes the trap amplitude at stationarity $L_s$, hence the stalling force $F_s=\kappa_T L_s$. 

For bundles of rigid non interacting filaments in ideal solution conditions, equilibrium statistical mechanics predicts \cite {Hill.81,PPCR.16}
\begin{equation}
F_s=N_f \frac{k_bT}{d} \ln{(\hat{\rho}_1)}.
\label{eq:Fs}
\end{equation}
For  bundles of semi-flexible filaments facing an optical trap load (as depicted in figure \ref{fig:sys}), equilibrium statistical mechanics applied to discrete wormlike chains (d-WLC) with step size $d$ predicts slightly larger values of the stalling force (1-2 percent)\cite{PPCR.16}. In order to avoid unbounded escaping filaments and remain in strict equilibrium conditions, a maximum value must be imposed to the grafted d-WLC contour length: the upper limit $z^*(L)d$ (precisely defined in the caption of figure \ref{fig:sys}) corresponds to a grafted filament with uniform curvature $1/L$ squeezed at normal incidence between two parallel planes separated by a distance $L$. Equilibrium properties in presence of this constraint are pertinent to an actin bundle in the optical trap as long as the probability of the onset of escaping is negligible over the observation time. In ref. \cite{PPCR.16} we have shown that the non escaping region is limited by
\begin{equation}
L<L_l=\sqrt{\frac{\ell_p d}{\ln{\hat{\rho}_1}}}
\label{eq:Ll}
\end{equation} 
as this condition ensures a negligible population of filament sizes at the boundary of the escaping regime, $z^*(L)$ (see caption of figure \ref{fig:sys}). 
In these conditions, the force exerted by a bundle of compressed flexible filaments in an optical trap set up at stalling results from the action of individual filaments which are continuously swapping between two states illustrated in the right panel of figure \ref{fig:sys}: either they do not touch the wall hence not contributing to the force (filament a in right panel of the figure) or they touch the wall and exert a $L$-dependent force $f(L)$ (filament b in the figure) which, assuming adiabatic conditions\footnote{fast microscopic intra-filament relaxation with respect to wall diffusion and F-actin self assembly kinetics, see appendix \ref{sec:char_time} for details}, and the weak compression behavior of the WLC model \cite{Frey.06}, has the form $f(L)\propto k_BT\ell_p/L^2$. 
This implies in turn that the bundle at equilibrium needs to recruit an average fraction $x \propto (L^2/\ell_p)$ of touching filaments to face the $L$-independent stalling force eq. (\ref{eq:Fs}), at the equilibrium trap amplitude $L_s$ corresponding to the chosen values of $\kappa_T$. We observe that the rigid filament case is recovered in the limit $\ell_p \rightarrow \infty$ where the wall will get impulsive forces from the most advanced filament with $f_b(L)\rightarrow\infty$ and $x\rightarrow 0$ as $\ell_p$ diverges.

This $L$-dependent load repartition for flexible filaments at equilibrium is expected to persist in non-equilibrium relaxation against arbitrary loads, and hence affect the load-velocity relationship. The aim of the present paper is to investigate the impact of filament flexibility on the bundle non-equilibrium dynamics by exploiting stochastic dynamics (SD) simulations. The model of living d-WLC used in this work is similar to the one in ref. \cite{PPCR.16} except that the contour length restriction is replaced by a realistic compressional response of the d-WLC in the strong compression regime, obtained by Monte Carlo calculations, which allows to make excursions into the escaping regime within a unique model. We follow the relaxation of a flexible bundle against a harmonic force mimicking the experimental conditions of ref. \cite{Footer.07}, as illustrated in the left panel of figure \ref{fig:sys}. The trap strength at given $\hat{\rho}_1$ and $N_f$ controls the range of the bundle length investigated, therefore the occurrence of escaping filaments. Similar to the rigid filaments case\cite{PPCR.18}, the wall relaxation is expected to be mostly adiabatic which allows us to consider the trap relaxation as a sequence of stationary non-equilibrium states of the bundle at different loads.
This implies that the celebrated "velocity-load" relationship $V(F)$, used to characterize the system behavior in rigid filament models, can be generalized to a "velocity-load-bundle length" relationship $V(F,L)$ without the need to consider complicated history-dependent behaviors. By tuning $\kappa_T$ in the optical trap set up, we can generate non-equilibrium steady-states for independent values of the load $F$ and the bundle length $L$ and characterize the relationship in the entire relevant range of the parameters.  

The bulk of the paper is organized as follows. Our dynamical model of living d-WLC \cite{PPCR.16} in an optical trap apparatus is described in section 2. In the following section, Stochastic Dynamics (SD) simulations are exploited to follow the relaxation of the bundle against an optical trap load, for a range of trap strengths. Peculiar features due to flexibility are emphasized by a direct comparison with the rigid filament predictions. A useful concept of chemical friction linked to the trap relaxation time is introduced in section \ref{sec:VFL}. 
Some results on the dynamical evolution towards the escaping regime are illustrated in section \ref{sec:tauOT}, while discussions and conclusions are provided in section \ref{sec:escaping}. Details of results and analytical derivations are collected in six appendices. 
To help the reader following the developments of our paper, we have regrouped in a separate table most physical quantities (mesoscopic or microscopic) appearing in our paper (see appendix \ref{app_table}).

\section{The dynamical model}
\label{sec:dynmodel}
We consider a grafted bundle of $N_f$ linear flexible filaments in supercritical conditions ($\hat{\rho}_1>1$) hitting a mobile hard wall which is subject to an external harmonic load as illustrated schematically in figure \ref{fig:sys}. 
The filaments are d-WLC with actin persistence length $\ell_p=14.5 \mu m$ and discrete step $d=2.7nm$ equal to the incremental contour length associated to the incorporation of a single actin monomer in the F-actin protofilament. 
Filaments are grafted normally to the wall located at $x=0$ by fixing permanently the position of the two first articulation points of the $n-th$ d-WLC filament ($n=1,N_f$) at $h_n$ and $h_n+d$ respectively along the $x-$axis (where $h_n$ is a longitudinal shift with $\left|h_n\right|<d/2$). 
In multi-filament bundles, two longitudinal arrangements are usually considered. 
In the unstaggered bundle case, the filament seeds are all located at the origin, $h_n=0$, while in the staggered bundle case, the filament seed positions are uniformly distributed within an amplitude $d$ according to
\begin{equation} 
h_n=\left[\frac{n-0.5}{N_f}-0.5\right] d \hskip 1cm n=1\dots N_f.
\label{eq:seeds}
\end{equation} 
The sizes of the filaments $\bar{j}=\{j_n\}_{n=1,N_f}$ are defined as the number of articulation points in each d-WLC filament. The minimum size $j_n=2$ represents the permanent oriented seed from which a filament can (re)grow.
\begin{figure}
\centering
\includegraphics[width=0.7\columnwidth]{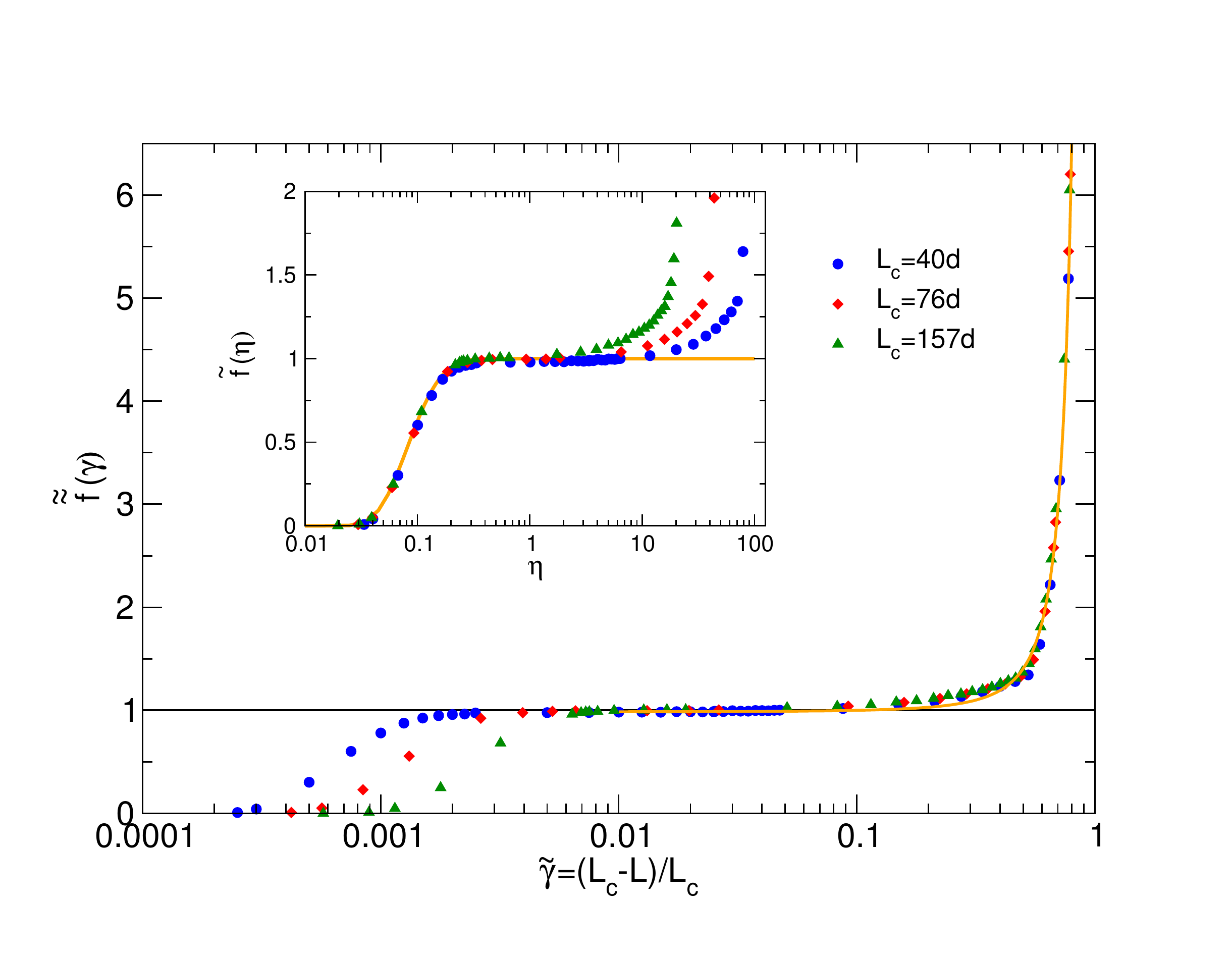}
\caption{Reduced force exerted by a passive d-WLC against a rigid wall located at $L$ as a function of $\widetilde{\gamma}=(L_c-L)/L_c$ for three different values of the contour length $L_c$. The force grows from zero to its buckling value (represented by the plateau) within a small range of $\widetilde{\gamma}$ depending on $L_c$ and departs from the unitary plateau at $\widetilde{\gamma}\simeq 0.1$. The orange solid line is the fitting function eq. (\ref{fit}). In the inset, the same reduced force as a function of $\eta=\widetilde{\gamma}\ell_p/L_c$. The orange solid line in inset is the weak bending universal curve eqs.(\ref{eq:frey}-\ref{eq:zetaeta}).
}
\label{fig:Fig1SM}
\end{figure}

We now specify the stochastic model for the coupled dynamics of the continuous variable $L$, the position of the moving obstacle, and of the $N_f$ discrete filament sizes.
The wall dynamics is described by the overdamped Langevin equation
\begin{equation}
\xi \frac{d L(t)}{dt}= F_{bun}(t)-\kappa_T L(t) + R(t)
\label{eq:Langevin} 
\end{equation}
where $\xi$ is the solvent friction coefficient on the obstacle related to its diffusion coefficient $D=k_BT/\xi$, where $T$ is the absolute temperature and $k_B$ the Boltzmann constant. 
The use of the overdamped limit is motivated by experimental conditions \cite{Footer.07,DCBB.14}, as detailed in appendix \ref{sec:char_time}. 
The second term in the rhs of eq. (\ref{eq:Langevin}) represents the action of the optical trap with a linear restoring force with trap force constant $\kappa_T$.
The fluctuation dissipation theorem requires that the white noise random force $R(t)$ satisfies
\begin{equation}
\langle R(t)\rangle=0\ \ \ \langle R(t+t') R(t')\rangle = 2 \xi k_BT \delta(t).
\label{eq:fluc} 
\end{equation}
On a coarse-grained time scale, $F_{bun}(t)$ represents the evolution of the total force exerted by the filaments on the wall. 
It depends on time through the wall position $L(t)$ and the filament sizes $\bar{j}(t)$ by expressing the bundle force as the sum of single filament forces
\begin{align}
F_{bun}(\bar{j},L)=\sum_{n=1}^{N_f} \bar{f}_{j_n}(L_n).
\label{eq:fbun}
\end{align}
where $L_n=L-h_n$ is the distance between the seed location of filament $n$ and the position of the obstacle $L$. 
$\bar{f}_{j_n}(L_n)$ is the equilibrium average force on a planar hard wall located at $L$, exerted by a passive (i.e. non-reacting) d-WLC of contour length $L_{cn}=(j_n-1)d$ with its two first articulation points clamped at $h_n$ and $h_n+d$. 
Such an adiabatic treatment of the individual filament forces on the moving wall is justified by the time scale hierarchy $\tau_{intra}\ll \tau_D \ll\tau_{chem}$ existing between the intramolecular relaxation time of the filament ($\tau_{intra}$), the local wall diffusion time $\tau_D=d^2/D$ and the characteristic time of the chemical reactions taken as $\tau_{chem}=W_0^{-1}$ (see appendix \ref{sec:char_time} for a detailed comparison of these time scales)
Direct filament-filament mutual interactions are disregarded.

For living filaments, i.e. filaments that undergo de/polymerization reactions, the total free energy of a clamped filament of size $j_n$ with seed-wall distance $L_n$ is
\begin{equation}
w_{j_n}(L_n,\hat{\rho}_1)=-k_BT \ln{\alpha_{j_n}(L_n)} -(j_n-2) k_BT \ln{\hat{\rho}_1}+C
\label{eq:fe}
\end{equation}
The first term, $-k_BT \ln{\alpha_{j_n}(L_n)}$, is the free energy penalty due to compression with $\alpha_{j_n}(L_n)$  the ratio of the partition functions of the clamped filament in presence and in absence of the wall \cite{PPCR.16}. 
The second term is the free energy gain due to the  self-assembly of $(j_n-2)$ monomers to the seed to form the filament \cite{Hill.81}. 
Finally, $C$ is an irrelevant, $j$--independent constant (the free energy of the permanent seed). 
The force $\bar{f}_{j_n}(L_n)$ in eq. (\ref{eq:fbun}) is thus
\begin{equation}
\bar{f}_{j_n}(L_n)=-\frac{\partial w_{j_n}(L_n,\hat{\rho}_1)}{\partial L_n}
=-\frac{k_BT}{\alpha_{j_n}(L_n)}\frac{\partial \alpha_{j_n}(L_n)}{\partial L_n}
\label{eq:force_i}
\end{equation} 

\begin{figure}
\begin{center}
\includegraphics[width=0.6\columnwidth]{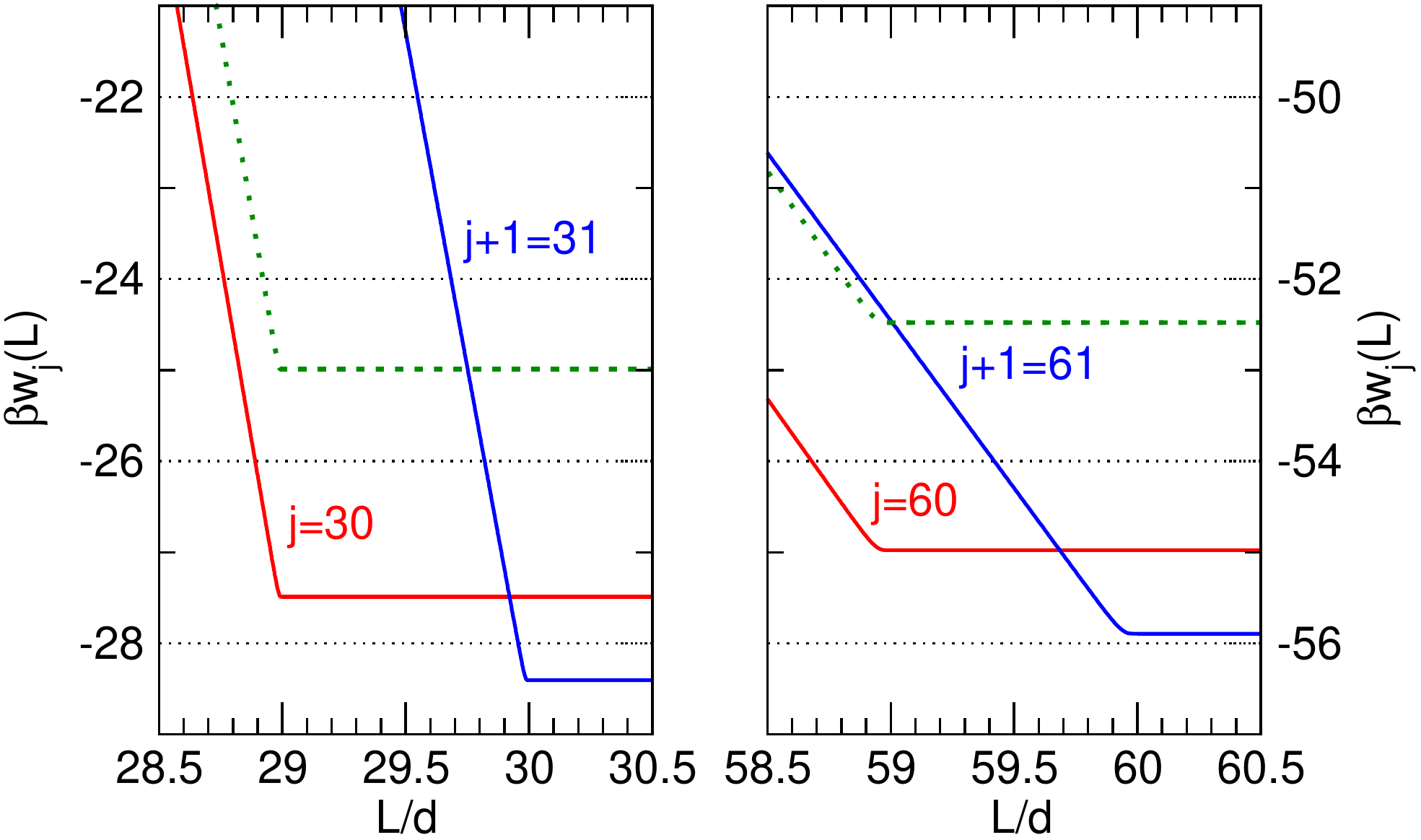}
\caption{Wall position ($L$) dependence of the reduced free energy $[\beta w_{j}(L)]$ of fixed-length grafted d-WLC filaments having neighbouring sizes $j$(red) and $j+1$(blue) for $j=30$ (left) and $j=60$ (right) ($L_c=(j-1)d$). The potential remains constant for the L regime where there is no direct contact between the wall and the fluctuating grafted d-WLC filament of size $j$, that is as long as $j \leq z(L)$, as defined and illustrated in figure \ref{fig:sys}. The vertical difference between blue and red profiles is the free energy gap in the exponential factor for the polymerization rate $U(j,L)$ at fixed wall position. The green-dotted profiles indicates the threshold of 2.5 between the significant and negligible  polymerization rate regions. Polymerization transitions from $30 \rightarrow 31$ are seen to be rare (less than $0.1 W_0$) if the wall is closer than $0.75d$ from the tip of an uncompressed filament of size $j=30$, hence for $L<29.75d$.
Conversely, polymerization transitions from $60 \rightarrow 61$ are possible ($\lesssim 0.1 W_0$) even if the filament of size $j=60$ is already under compression.} 
\label{fig:fig6}
\end{center}
\end{figure}

The form of $\bar{f}_j(L)$ 
for a passive d-WLC of persistence length $\ell_p$, contour length $L_{c}=(j-1)d$, clamped at $h=0$ and hitting a hard wall located at $L$ and normal to the grafting direction, is given by 
\begin{equation}
\bar{f}_j(L | L_c,\ell_p)=f_b(L_c,\ell_p) \times
\begin{dcases}
\tilde{f}\left(\widetilde{\gamma} \frac{\ell_p}{L_c}\right) \qquad &\widetilde{\gamma} < 0.1\\
\tilde{\tilde{f}}(\widetilde{\gamma}) \qquad&\widetilde{\gamma}\geq 0.1
\end{dcases}
\label{eq:force}
\end{equation}
where $\widetilde{\gamma}=(L_c-L)/L_c$ is the relative compression and $f_b$ is the Euler buckling force for a clamped rod \cite{b.Howard}
\begin{equation}
f_b(L_{c},\ell_p)=\frac{\pi^2}{4} \frac{k_B T \ell_p}{L_{c}^2}.
\label{eq:buckling}
\end{equation}
The form of the functions $\tilde{f}(\eta)$ and $\tilde{\tilde{f}}(\widetilde{\gamma})$, where $\eta=\widetilde{\gamma} \frac{\ell_p}{L_c}$, is illustrated in figure \ref{fig:Fig1SM} and discussed in details in appendix \ref{sec:WLC}. 
At weak bending ($\widetilde{\gamma} < 0.1$), $\tilde{f}(\eta)$ rises from zero to one over a short compression range $(L_{c}-L)\approx 0.25 L_{c}^2/l_p$ (which is typically less than $d$ for actin filaments up to $j \approx 150$)  (see the inset in figure \ref{fig:Fig1SM}), and remains constant over a large range of compressions up to $\widetilde{\gamma} \simeq 0.1$. For larger compression a universal behavior appears in terms of $\widetilde{\gamma}$ and has been characterized numerically as detailed in appendix \ref{sec:WLC}. 
The compression law at weak bending, derived analytically in ref. \cite{Frey.06}, leads to a linear decay of the free energy $w_{j}(L,\hat{\rho}_1)$ with $L$, which extends almost up to $L=L_c$ beyond which the filament force vanishes as illustrated in figure \ref{fig:fig6}.\footnote{The change in slope is just a little before $L=L_c$ because some small compression is needed to reach the buckling regime, see figure \ref{fig:Fig1SM}.}

Like in Brownian Ratchet models \cite{PPCR.18}, the stochastic time evolution of the filament sizes $\bar{j}(t)=\{j_1(t),\dots,j_n(t),\dots,j_{N_f}(t)\}$ is specified by a Poisson distribution of waiting times for the next instantaneous single monomer polymerization/depolymerization. 
For each filament $n$ ($n=1,N_f$), we adopt rates 
\begin{eqnarray}
U_n(j_n,L)&=&U_0 \frac{\alpha_{j_n+1}(L_n)}{\alpha_{j_n}(L_n)}\nonumber\\
&=&W_0 \frac{\exp\{-\beta w_{j_n+1}(L_n,\hat{\rho}_1)\}}{\exp\{-\beta w_{j_n}(L_n,\hat{\rho}_1)\}}\;\; \text{if}\  j_n \geq 2 \label{eq:U}\\
W_n(j_n,L)&=&\begin{cases}W_0  & \text{if}\  j_n >2  \\   0  & \text{if}\  j_n =2\end{cases}
\label{eq:W}
\end{eqnarray}
similar to the ones for the rigid filaments case given in ref. \cite{PPCR.18}. For rigid filaments ($\ell_p\rightarrow\infty$) the $\alpha$ factors are either 1, if the filament does not interact with the wall, or 0 in case the filament touches the wall. 
For the flexible case, we still have $\alpha_{j_n}(L_n)=1$ for $j_n < z(L_n)$, where $z(L_n)=\text{int}(L_n/d)+1$ (see figure \ref{fig:sys}) is the largest size of a filament such that, for a given seed-wall distance $L_n$, none of its microscopic configurations overlaps with the wall. 
Given the conditional equilibrium size distribution $P_{j}(L,\hat{\rho}_1) \propto \exp{(-\beta w_{j}(L,\hat{\rho}_1))}$ \cite{PPCR.16}, it is readily verified that the adopted rates given by eqs.(\ref{eq:U}),(\ref{eq:W}) satisfy micro-reversibility at equilibrium. 
Our choice assumes that the depolymerization rate $W_0$ of the filament is independent of the G-actin monomer concentration and is insensitive to any form of filament-wall contact. 
The effective polymerization rate, equal to $U_0^{c}=k^{on} \rho_{1c}=W_0$ at the critical monomer concentration $\rho_{1c}$ in the bulk, increases linearly with the bulk monomer concentration but decreases with the work needed to compress the filament as its length is increased by $d$ at fixed $L$. 
This is illustrated in figure \ref{fig:fig6} where the $L$ dependence of two free energy expressions $w_{j}(L,\hat{\rho}_1)$ and $w_{j+1}(L,\hat{\rho}_1)$ for successive filament sizes is shown respectively for $j=30$ and $j=60$. 
The observed linear increase of $w_{j}(L,\hat{\rho}_1)$ as the wall position decreases from $L=L_{cj}=(j-1)d$ is a direct consequence of the constant compression force at weak bending, see figure \ref{fig:Fig1SM}. 
The difference between the horizontal plateaus for two successive sizes is the free energy change when adding/removing one monomer, $-k_BT\ln \hat\rho_1$. 
If we arbitrarily consider $e^{-(\beta \Delta w)} W_0=e^{-2.5} W_0 \approx 0.1 W_0$ as the crossover rate between ``negligible" and ``significant" polymerization rates, figure \ref{fig:fig6} shows that the polymerization $j\rightarrow j+1$ transition rates are significant for a filament $j=60$ even when it is already under compression. 
By contrast, for a shorter filament of size $j=30$, polymerization rates of compressed filaments are negligible and become significant only when the wall lies beyond $0.75d$ from the filament tip, i.e. in the region where the blue profile lays below the green dotted profile. 

The time evolution of our model of coupled wall-filaments dynamics has been generated by the Explicit-Wall algorithm (EWA) proposed in  \cite{GG.10,GG.07,Ca.08,PPCR.18} which treats both the filaments sizes and the wall position as discrete random variables to reformulate the time evolution as a purely Markov chain dynamics. 
This step requires to replace the Langevin equation for the wall dynamics by an equivalent Fokker-Planck equation (see appendix \ref{sec:EWA}). 
To make our simulations feasible, we adopted the value $\epsilon=\tau_D/\tau_{chem}=W_0d^2/D=0.0469$ (where $D$ is the wall diffusion constant) which in ref. \cite{PPCR.18} was shown to accurately represent the experimentally relevant situation of quasi infinite wall diffusion ($\epsilon_{exp}=4.69 \times 10^{-5}$).  

Before presenting the result of our non-equilibrium simulations, we would like to discuss the nature of the steady and the equilibrium states of our model. 
We follow the relaxation of an initially short bundle under the action of polymerization and trap restoring forces. 
In supercritical conditions the trap amplitude grows until the reach of a steady state.
For rigid filaments this is a genuine equilibrium state where the pushing polymerization force is on average balanced by the trap force and the trap amplitude is inversely proportional to the trap strength \cite{PPCR.18}. 
Also for semiflexible filaments we observe the reach of a steady trap amplitude at large times. 
However, this corresponds to a metastable local equilibrium state, where polymerization force is balanced by the trap load, only for trap strength large enough to have $L_s<L_l$ as given in eq.(\ref{eq:Ll}). 
If this condition does not hold, we incur in the escaping regime where the steady trap amplitude does not correspond to a force balancing the polymerization force, as detailed in section \ref{sec:escaping}. 
In the latter case we don't have an equilibrium state since filament lengths grow indefinitely. 
In ref. \cite{PPCR.16} we have carefully characterized the stability of the non-escaping local equilibrium state. 
Being a metastable state, there is always the possibility that even in the non-escaping regime, one filament becomes escaping by running trajectories for long enough time. 
However, the probability of filaments with contour length longer than $L$ decreases exponentially fast with $L_c-L$ at fixed $\kappa_T$, i.e. the probability per unit time to observe the occurence of an escaping filament becomes exponentially small\cite{BM.05}.

We point out that while the control of the $L$ fluctuations is particularly easy in an optical trap device, specifically by adjusting the trap force constant, the $L$ fluctuations and the associated onset of escaping filaments are, on the contrary, much more difficult to predict in constant force experiments where no local equilibrium exists \cite{BM.05,BM.06}.

\section{Relaxation of a staggered semi-flexible bundle in an optical trap}
\label{sec:result}
We have studied systems at several values of $N_f$ and $\hat\rho_1$ with both staggered and unstaggered seed dispositions, but here we first concentrate on the specific case of a staggered bundle of $N_f=32$ filaments with seeds disposition as in Eq.\eqref{eq:seeds} at $\hat{\rho}_1=2.5$. In the following for convenience, the staggered rigid bundle model (in the infinite diffusion limit) will be specifically referred to as SR (for staggered rigid).
\begin{figure}[!h]
\includegraphics[width=0.5\columnwidth]{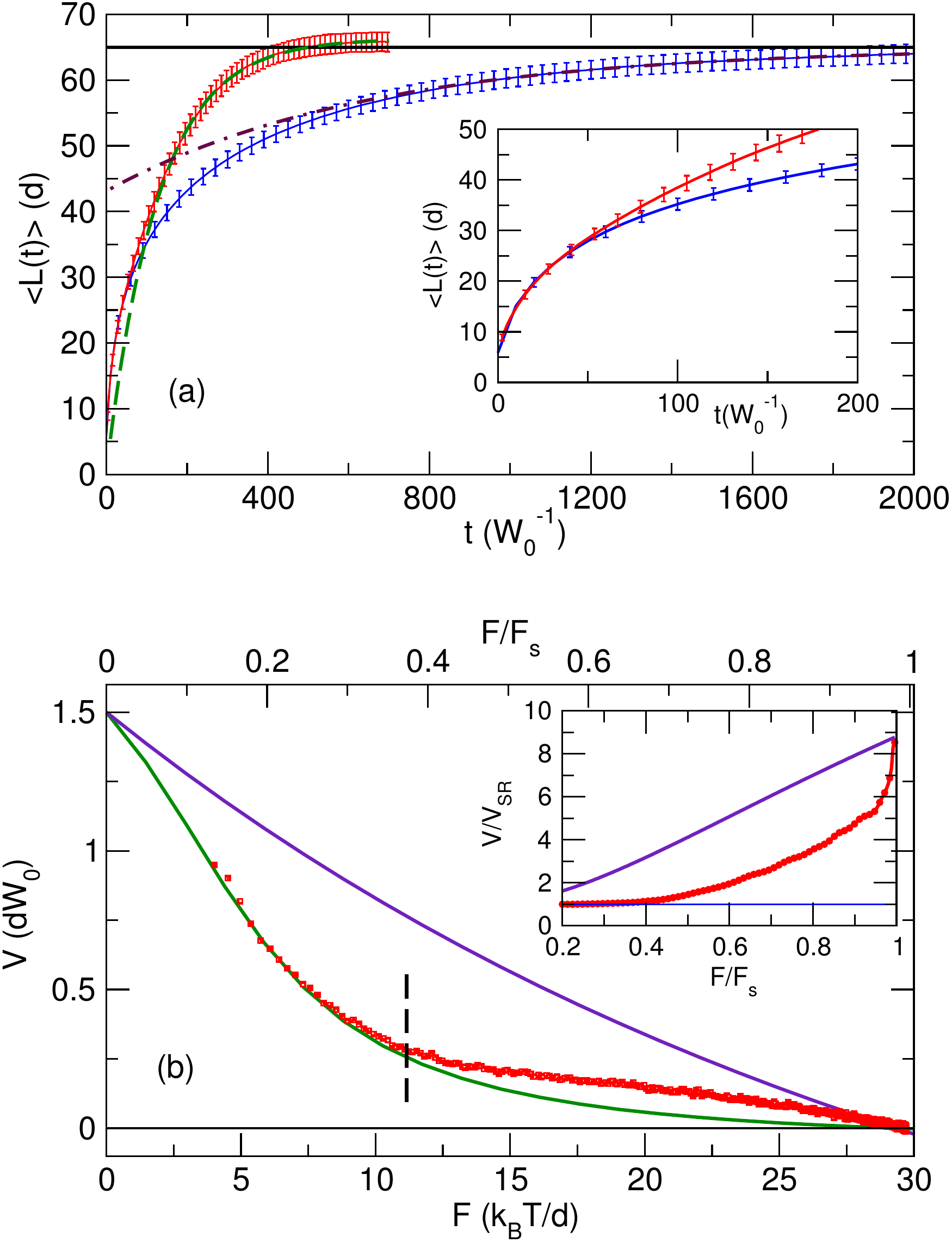}
\caption{\small Panel (a): Relaxation of a staggered bundle of $N_f=32$ filaments subject to optical trap load with trap strength $\kappa_T=0.4511k_BT/d^2$ at $\hat{\rho}_1=2.5$. Results for rigid (blue line) and flexible (red line) models are reported.
The horizontal line represents the value $L_s=F_s/\kappa_T=65d$ from eq. (\ref{eq:Fs}), while the flexible model has the slightly larger asymptotic value, in agreement with the theoretical prediction $L_s=66.1877d$ \cite{PPCR.16}. 
The vertical bars on the data indicate the standard deviation of $L(t)$ which is seen to evolve rapidly towards its predicted equilibrium values $\sigma=(k_BT/\kappa_T)^{1/2}=1.489d$ and $\sigma=1.5178d$ for the rigid and the flexible models respectively. Exponential fits to the long time behavior for both models are shown by dashed and dot-dashed lines. 
Numerical values of the parameters are reported in the text. The inset is an enlargement of the region where flexible and rigid model behaviors start diverging from each other.\\
Panel (b): $V(t)=d\langle L(t)\rangle/dt$ versus $F(t)=\kappa_T \langle L(t)\rangle$ for both rigid (green line) and flexible (red squares) models. The green line is an accurate theoretical prediction of the stationary velocity for the SR model \cite{DCBB.14}.
The purple curve represents the PLS behavior from eq. (\ref{eq:PLS}). Inset: velocity relative to the SR prediction  versus the fractional load: PLS (purple line), flexible model (red points).}
\label{fig:fig2}
\end{figure}
In order to highlight the effects of flexibility we compare results for flexible and rigid bundles at the same thermodynamic conditions in figure \ref{fig:fig2}. 
The top panel shows the average wall relaxation $\langle L(t)\rangle$ of both a rigid bundle ($\ell_p=\infty$) and a bundle with the F-actin value $\ell_p=5370d$ in an optical trap at $\kappa_T=0.4511 k_BT/d^2$, a value large enough to avoid escaping filaments for the flexible case \cite{PCR.15,PPCR.16}, but weak enough to allow flexibility effects to manifest.  
Results for the rigid model have been already discussed in \cite{PPCR.18}. 
For the flexible bundle, we followed, for a time interval of $T=700 W_0^{-1}$, the relaxation of a statistical sample of 10$^3$ replica with initial configurations sampled by the equilibrium size distribution with the wall fixed at $L(0)=5d$ \cite{PPCR.16}. 
The observed asymptotic values of the wall position and its variance are in full agreement with the equilibrium theory predictions (see the  caption of figure \ref{fig:fig2}) \cite{PPCR.16}. 
The short vertical bars on the data represent the standard deviation $\sigma$, not the statistical error. 
Its rather limited value at all times is a unique characteristic of the optical trap apparatus which allows to obtain rather precise measurements already with a limited number of relaxations (just one in real experiments! \cite{Footer.07}). 

After a fast initial relaxation, $\langle L(t)\rangle$ approaches exponentially its asymptotic value $L_s$  
\begin{equation}
\label{eq:Lt}\langle L(t)\rangle-L_s \sim -A e^{-t/\tau^{OT}}
\end{equation}
where $\tau^{OT}$ is the final relaxation time and $A$ the amplitude of the slowest relaxation mode which depends on the initial conditions. 
Fits, represented in the figure by dashes lines, provide $A=65.9 d$, $\tau^{OT}=128 W_0^{-1}$ and $A=21.8 d$, $\tau^{OT}=653 W_0^{-1}$ for flexible and rigid bundles respectively.
Velocity and load both follow analogous long time behaviors with the same relaxation time which can thus be expressed as 
\begin{equation}
\tau^{OT}=\frac{1}{\kappa_T} \lim_{t \rightarrow \infty}  \frac{F_s-F(t)}{V(t)}
\label{eq:tauOT}
\end{equation}
where $V(t)=d\langle L(t)\rangle/dt$ and $F(t)=\kappa_T\langle L(t)\rangle$ is the time dependent load.
Note that $\tau^{OT}\gg \tau_{micro}\sim \tau_{chem}=W_0^{-1}$ in both models, where $\tau_{micro}$ is the relaxation time of the distribution of the filament tips with respect to the wall position. 
More precisely, it was shown in the rigid bundle case \cite{PPCR.18} that $\tau_{micro}$ is of the order of few $W_0^{-1}$ and the distribution of filament tips is peaked at the wall position with a characteristic width of the order of unity, as long as $F/F_s>0.15$. 
In the long time relaxation, for both the rigid and the present flexible bundles, the load changes very slowly on the time interval over which the chemical events take place and the bundle tips relax. 
This separation of time scales implies that during the slow variation of the trap amplitude, and thus of the load, the distribution of the filament lengths (with respect to the position of the obstacle) remains equivalent to the distribution in a hypothetical constant load experiment with load value set to $\kappa_T \langle L(t)\rangle$ at the current stage of the relaxation process.\footnote{As $F/F_s \rightarrow 0$, the separation of time scales brakes down as both $\tau_{micro}$ and the width of the size distribution diverge \cite{PPCR.18,DCBB.14}.} 
This finding holds for any value of $N_f$ and $\kappa_T$ in the non-escaping regime which emphasizes the general character of our results. 
 
The bottom panel of figure \ref{fig:fig2} reports the wall velocity $V(t)$, obtained by numerical differentiation, and plotted as a function of the time dependent load $F(t)$ for both rigid and flexible bundles. 
Rigid model behavior is represented by the approximate solution of ref. \cite{DCBB.14} which turns out to be very accurate (see also ref. \cite{PPCR.18} for the accuracy of this result).
On the same plot we also report for reference the so called ``Perfect Load Sharing'' (PLS) behavior given by 
\begin{eqnarray}
V&=&d \left[U_0 \exp{\left(-\frac{dF}{N_f k_BT}\right)}-W_0\right]
\label{eq:PLS} 
\end{eqnarray}
which implies an equal repartition of the work among all filaments\cite{Borizy.08,Sander.20}. Equivalently, eq. (\ref{eq:PLS}) expresses that the power of transduction of chemical energy into mechanical work, $P=V~F$, results to be $N_f$ times the power developed by a single filament against a force $f=\frac{F}{N_f}$. Such equal repartition is interesting as it was shown to maximize the power of transduction $P=V~F$ \cite{Borizy.08,WC.14}. 
The $V(F)$ data for the flexible model at the chosen $\kappa_T$ value closely follow the rigid bundle curve up to $F\approx(12-13)k_BT/d$ (i.e. $F/F_s\approx 0.35$, indicated by the vertical bar in the bottom panel of the figure), corresponding to $L \approx 25d$ at that $\kappa_T$ value (see also the inset of the top panel, where the velocity starts deviating significantly to larger and larger values). As stalling is approached, the $V(F)$ data get closer to the PLS curve (see the inset in the bottom panel).
This behavior shows that the velocity for flexible bundles is function not only of the external load but also of the bundle length.  

The observed strong reduction of $\tau^{OT}$ for flexible bundles is a manifestation of a new polymerization mechanism linked to flexibility, a mechanism first suggested by Mogilner and Oster in a series of seminal papers \cite{Mogilner.96,MO1996b,MO2003} where the Elastic Brownian Ratchet (EBR) model was introduced. This mechanism improves the work repartition over the filaments as the fraction of filaments simultaneously touching the wall correlates with the load sharing capacities of the filaments in the bundle. 
When filaments are short, the buckling force $f_b(L)$ they develop when in contact with the wall is rather large and few filaments (just a single one acting sporadically for very tight traps) are able to sustain the entire load. 
Therefore in a very short trap the dynamic behavior of the flexible bundle is equivalent to the rigid model behavior, as seen in the inset of figure \ref{fig:fig2}(a) for $L<(25-30)d$.
When filaments get longer, the single filament buckling force becomes increasingly weaker and a significant fraction of filaments has to be recruited simultaneously to sustain the increasingly large external load. 
As $L$ further increases, all filaments would be recruited to act permanently and the PLS picture would become valid. This situation however is never reached in the present model because thermal bending fluctuations cause filaments to escape before the PLS is reached, as discussed in section \ref{sec:escaping}.
In eq. (\ref{eq:tauOT}) we have defined the optical trap relaxation time, an experimentally accessible quantity, and found that it may shorten significantly for flexible bundles as a result of improved load sharing capacities. We come back on this important and original issue in the next section where $\tau^{OT}$ will be related to the trap strength and to an effective chemical friction term reflecting precisely the flexibility influence on the bundle dynamics close to stalling.

\section{Analysis of the filament flexibility on the load-velocity relationship.}
\label{sec:VFL}
The increase of velocity induced by flexibility originates from the additional polymerization mechanism in which the monomer addition inside a tip-wall gap smaller than $d$ is made possible by a bending fluctuation. 
To interpret quantitatively the onset of this mechanism in the appropriate high wall diffusion limit, we first reconsider the SR model. 
The fast wall motion is blocked by the most advanced tip in the bundle. 
Consider a second filament which, because of the seeds shift, has its tip located at a distance $n\delta$ with $n\in [1,N_f-1]$ from the tip of the most advanced filament. 
If the wall position fluctuates beyond $(N_f-n) \delta$ away from the most advanced tip, the gap between the tip of the second filament and the wall is $\geq d$ and this filament can polymerize with a rate $U_0$. 
This is the subsidiary mechanism of the SR brownian ratchet bundle which enhances the wall velocity with respect to the unstaggered rigid bundle case and hence improves the load sharing capacities of the bundle just as a result of a more favorable disposition of the seeds of the rigid filaments \cite{WC.14}.

\begin{figure}
\includegraphics[width=0.4\columnwidth]{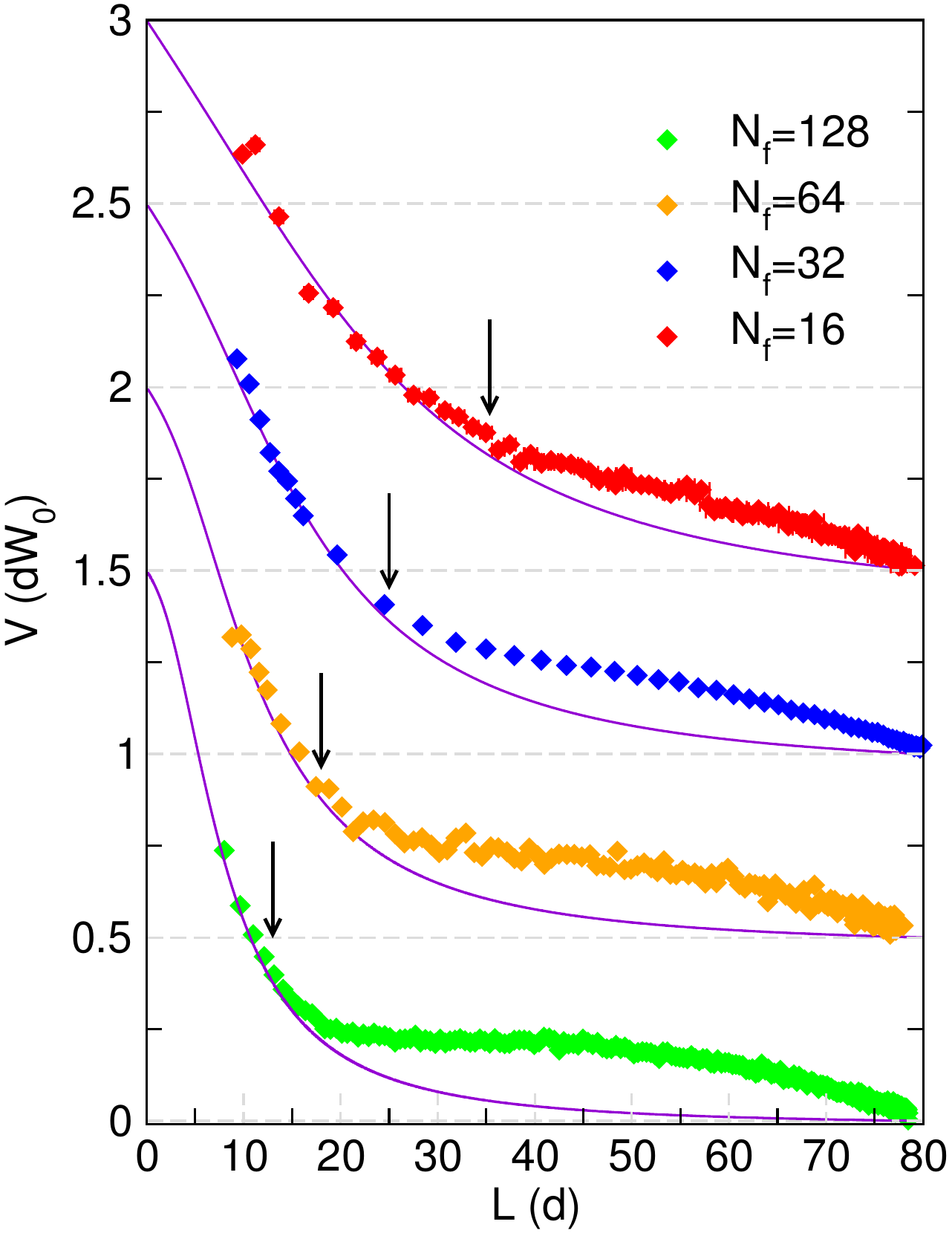}
\caption{Flexible bundles relaxation experiments against a harmonic load with trap strength $\kappa_T=0.011453\; N_f~k_BT/d^2$ chosen to cover a similar $L$ window between the bundle initial size $L_0\approx 5d$ up to a common equilibrium value $L_s\approx 80 d$. The plots show the instantaneous average wall velocity as a function of the trap width. The continuous lines show, for reference, the SR behavior in identical conditions based on D{\'e}moulin et al. theory \cite{DCBB.14}. The data points are the average over $128$ (for $N_f=64,128$), $256$ (for $N_f=16$) or $992$ (for $N_f=32$) independent trajectories. The crossover to the escaping regime is $L_l=76.6d$ at $\hat{\rho}_1=2.5$ from eq. (\ref{eq:Ll}) but trajectories are stopped before observing any escaping filament. The flexibility effect is seen to start at a distance close to $\Lambda$ given by eq. (\ref{eq:Lstar}) and indicated by the vertical arrows. For sake of clarity the vertical axis of different experiments are shifted upward by 0.5 $dW_0$.} 
\label{fig:flex}
\end{figure}

In the same conditions, but now with flexible filaments, the above mechanism exploiting wall fluctuations will remain operative. 
In addition, as the wall fluctuates around its most probable position directly against the most advanced tip, the second filament in the above example could alternatively polymerize by undergoing a $n\delta$ compression, resulting into another kind of subsidy effect originally suggested for the $13$ protofilaments microtubule facing a non fluctuating wall \cite{Mogilner.99}. 
Such a transition towards a compressed grafted filament should be followed by a relaxation of the filament by displacing the wall against the external force.
A compression by $n\delta$ plus the addition of a monomer becomes energetically possible when the difference in free energy between the uncompressed state and the compressed state with one added monomer satisfies 
$n\delta f_{b}(L)-k_BT \ln{\hat{\rho}_1} \sim k_BT$, where $n\delta f_b(L)$ is the compression free energy difference and $-k_BT \ln\hat\rho_1$ is the free energy gain per monomer addition.\footnote{Here we replace the filament contour length by the trap length since they differ at most by few  $d$'s, close enough for our qualitative consideration \cite{Frey.06,PPCR.16}. }
This relation, together with eqs. (\ref{eq:force})-(\ref{eq:buckling}) at weak compression in the buckling state, gives a $n$-dependent effective length 
\begin{equation}
\label{eq:lbar}
\bar{L}(n) \approx \sqrt{\frac{n \pi^2 \ell_p d}{4 N_f (1+\ln{\hat{\rho}_1})}} 
\end{equation}
for a bundle of $N_f$ filaments at supercriticality $\hat\rho_1$. 
In panel (a) of figure \ref{fig:fig2}, the flexibility effects for a $N_f=32$ bundle with $\ell_p=5370d$ at $\hat\rho_1=2.5$ are observed to start at $25d \approx \bar{L}(3)$ (see the insets), corresponding, for the used trap strength $\kappa_T=0.4511 k_BT/d^2$, to $F\simeq 12.5k_BT/d$ indicated by a vertical dashed bar in panel (b) of the same figure. Further, a series of four additional similar relaxation experiments is presented in figure \ref{fig:flex} for four different bundle sizes $N_f=16,32,64,128$ in identical supercritical conditions $\hat\rho_1=2.5$. The trap strength has been set at $\kappa_T=0.011453\; N_f~k_BT/d^2$ in order to reach roughly the same stalling distance $L_s\approx 80 d$ at large time and to cover the same $L$ window from the initial value $\approx 5d$. We empirically observe that in all cases $\bar{L}(3)$, indicated by a vertical  arrow in the figure, signals the beginning of the flexibility influence. Hence, we adopt
\begin{equation}
\Lambda=\bar{L}(3)=\sqrt{\frac{3\pi^2 \ell_p d}{4 N_f(1+\ln{\hat{\rho}_1})}}
\label{eq:Lstar}
\end{equation}
as the crossover size between the rigid regime and the flexible regime for a staggered bundle of $N_f$ filaments. We note that the influence of $\hat{\rho}_1$ on $\Lambda$ (not studied here as all experiments are performed at the same supercritical conditions) is predicted to be marginal by comparison to the influence of $N_f$ or $\ell_p$ variables.

As the above analysis shows, the velocity enhancement due to filament flexibility depends on $L$. Therefore the velocity-load relation $V(F;N_f,\hat{\rho}_1)$  characterizing the behavior of SR models should be generalized to staggered semi-flexible bundles (in the experimentally relevant adiabatic trap regime) by including the bundle size $L$ as a relevant independent macroscopic variable, leading to the form $V(F,L;N_f,\hat{\rho}_1)$. 
Flexibility being more adequately represented by the reduced size $\lambda=L/\Lambda$, the alternative but equivalent form $V(F,\lambda; N_f,\hat{\rho}_1)$ can be used to describe the dynamical behavior of flexible bundles, the rigid case being recovered  for $\lambda \leq 1$.

\begin{figure}[ht]
\begin{center}
\includegraphics[width=0.7\columnwidth]{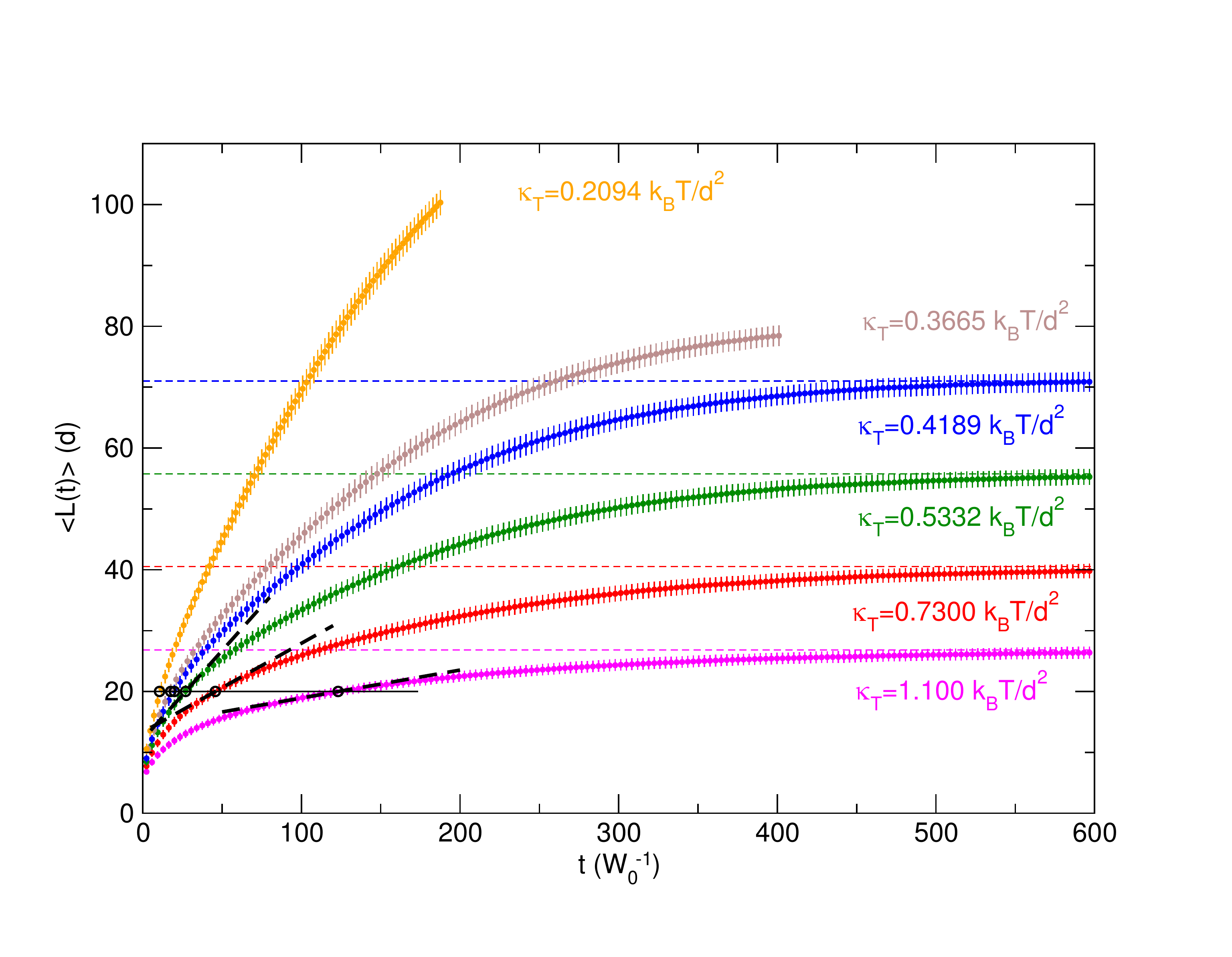}
\caption{Non-equilibrium dynamics of a staggered bundle of $N_f=32$ flexible filaments at $\hat{\rho}_1=2.5$ growing against an optical trap load. Time relaxation of the wall position $\langle L(t)\rangle$ and associated standard deviation $\sigma(t)$ at the indicated value of the trap strength $\kappa_T$. Data are obtained averaging the relaxation of $992$ equivalent replica with initial conditions sampled by the equilibrium distribution at $L(0)=5d$. As long as the plateau remains below $\approx 65d$ (see text), escaping filaments influence is absent and each relaxation curve goes asymptotically towards a plateau value $F_s/\kappa_T$ in full agreement with the equilibrium statistical mechanics prediction of the stalling force $F_s$ for the same model \cite{PPCR.16}, indicated by a horizontal dashed line. We observe that such an agreement is still verified for $\kappa_T=0.4189k_BT/d^2$ ($ L_s\approx 70d$) despite the presence of some escaping filaments. In all curve portions shown, we checked that the bias in the mean $\langle L(t)\rangle$ resulting from including or not data from trajectories with at least one escaping filament, did not exceed $0.03d$. The points where the different $\langle L(t)\rangle$ curves intersect an horizontal line (\emph{e.g.} the continuous black line at $L=20d$) lead to the fixed $L$ series of $(V,F)$ data reported in figure \ref{fig:fig1}. Along the constant $L$ horizontal line, the value of the force is given by $\kappa_T L$ while the value of the velocity by the local slope indicated as black dashed lines.} 
\label{fig:fig4}
\end{center}
\end{figure}

\begin{figure}[t]
\centering
\includegraphics[width=0.7\columnwidth]{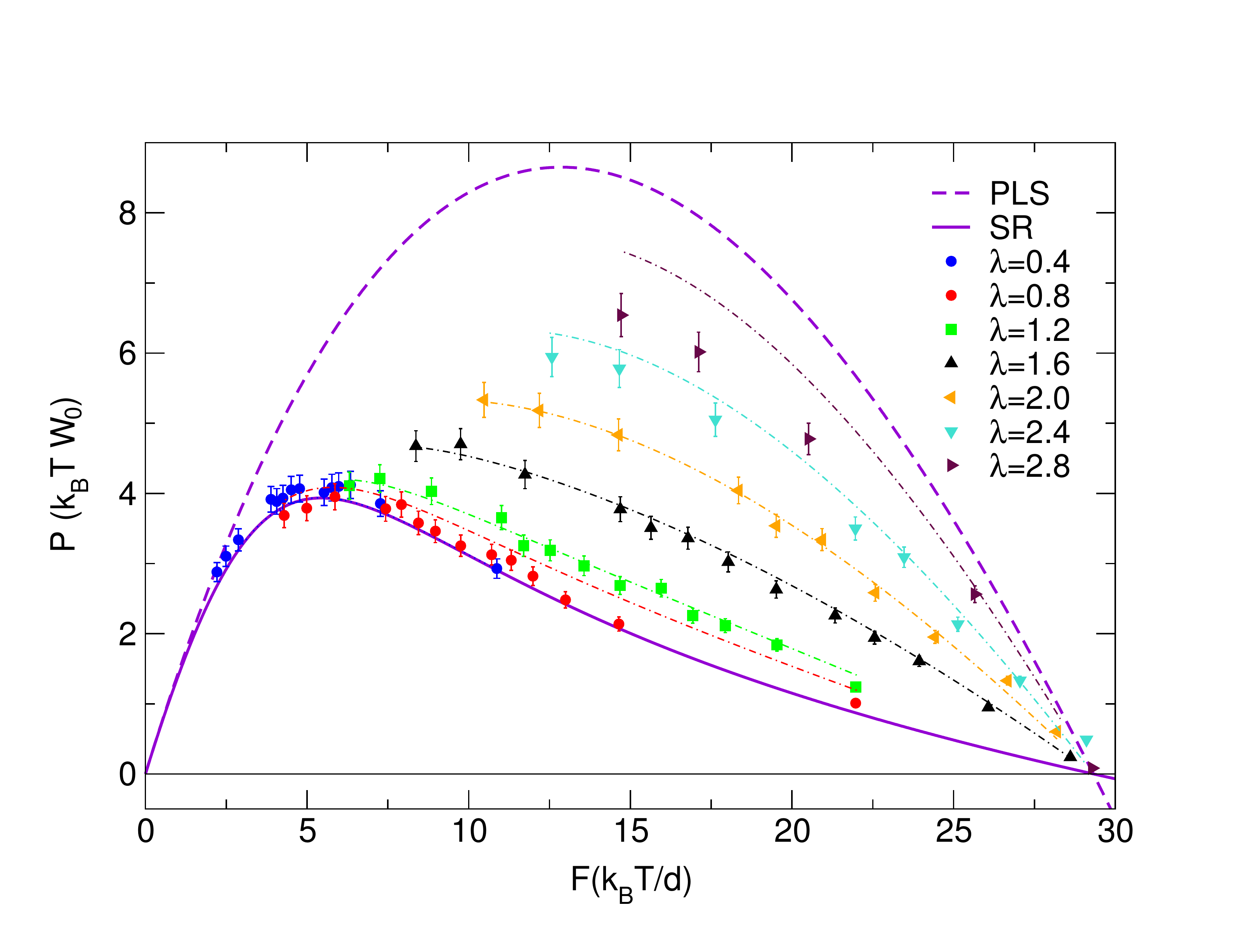}
\caption{Power $P(F,\lambda)=F V(F,\lambda)$ developed by the polymerization force of a homogeneous bundle of $N_f=32$ actin semi-flexible filaments, pressing against a mobile wall located at $L$ and subject to a load $F$ in super-critical conditions at $\hat{\rho}_1=2.5$. 
$\lambda=L/\Lambda$ measures the flexibility, where $\Lambda$ is the threshold distance between the rigid and flexible bundle domains defined in eq. (\ref{eq:Lstar}). 
Two theoretical force velocity predictions are also shown, the PLS prediction eq. (\ref{eq:PLS}) and the SR model prediction as represented by the D{\'e}moulin et al. approximation \cite{DCBB.14,PPCR.18}. At given load, a systematic increase of the power with $\lambda$ is observed, the two theoretical curves representing upper and lower boundaries to flexible data points.  Except at large $\lambda > 2.4$, data at fixed $\lambda$ can be fairly well represented by a linear combination of those boundaries with weights proportional to $\lambda^2$ (see eq. (\ref{eq:power})) shown by dot-dashed lines.} 
\label{fig:fig1}
\end{figure}

In order to map the entire $V(F,\lambda)$ relationship for the flexible bundle at a given $(N_f=32,\hat{\rho}_1=2.5)$ state point, we performed 14 different optical trap experiments (averages over $992$ individual trajectories in each case) at different trap strengths in the range $0.2094k_BT/d^2 \leq \kappa_T\leq 1.10 k_BT/d^2$. 
A subset of those dynamical relaxations is reported in figure \ref{fig:fig4}. 
The data analysis requires a careful treatment of trajectories affected by escaping filaments. 
While $L_l=76.6d$ obtained by eq. (\ref{eq:Ll}) at $\hat{\rho}_1=2.5$ is a good qualitative estimate of the crossover length to the escaping regime, a more quantitative threshold to the escaping regime in an optical trap at fixed $\kappa_T$, as derived in ref. \cite{PPCR.16} at stalling conditions is $ L_s \leq 65 d$ and $\hat{\rho}_1=2.5$.
This condition remains valid for the SD sampling of the bundle growth relaxation towards a final equilibrium state and it roughly corresponds to $\kappa_T^{min}\simeq 0.42 k_BT/d^2$ for the present system.
For the few lowest values of $\kappa_T <\kappa_T^{min}$ employed, some filaments in all dynamical replica were found to escape at some stage and in figure \ref{fig:fig4} we just show the portion of the average relaxation before the manifestation of the phenomenon. 
Approaching the threshold values of $\kappa_T$ from the small $\kappa_T$ side, a larger and larger fraction of replica doesn't present escaping filaments over the time window of the experiment. 
In these cases, this fraction is used for the statistical analysis. 
The resulting dynamical behaviors are then used to extract values of the velocity $V(F,\lambda)$.
Similar to the analysis performed to obtain the bottom panel of figure \ref{fig:fig2}, we obtain the velocity $V(t,\kappa_T)$ for each relaxation experiment by numerical differentiation and we eliminate the time in favor of the force $F(t)=\kappa_T \langle L(t)\rangle$. 
Since the bundle flexibility is directly related to its size, to interpret and rationalize the results in terms of the flexibility we group together data from the different experiments but at the same trap amplitude $L$. 
The justification of this procedure is again related to the observed separation of time scales between the slow load variation and the faster relaxation of filament lengths. 
The resulting picture is reported in figure \ref{fig:fig1} where we show the power of transduction of chemical into mechanical energy. 
For short filaments ($\lambda\leq 1$) the effect of flexibility is negligible and the power closely follows the SR behavior (partial load sharing \cite{Borizy.08}). 
As the bundle length increases ($\lambda > 1$), the power progressively increases towards the PLS mean field behavior of eq. (\ref{eq:PLS}) indicating that flexibility induces better work sharing capacities, as anticipated by Schaus and Borizy \cite{Borizy.08}. 
The PLS behavior, however, cannot be reached without entering the escaping regime occurring at $\lambda = L_l/\Lambda \simeq 3$ for the present system (see figure \ref{fig:fig3}). 

Data for the force dependence of the velocity or the power at fixed $L$ (or fixed $\lambda$) can be quantitatively represented by a linear combination of PLS and SR behaviors. We suggest the form 
\begin{equation}
V(F,L)\simeq b(L) V^{PLS}(F) + [1-b(L)] V^{SR}(F)
\label{eq:velcomb}
\end{equation}
with the simple \textit{ansatz} $b(L)=(L/L_l)^2$. 
Such a choice is suggested by the $L^2$ growth of the number of filaments recruited to produce a given force $F$ at different trap widths, as mentioned earlier in this section. 
The crossover length $L_l$, defined in eq (\ref{eq:Ll}) as the length where the bundle enters massively in the escaping regime, is here interpreted as the upper limit of the non escaping regime where all filaments are permanently touching the wall, hence leading to the PLS behavior.
In figure \ref{fig:fig1}, eq. (\ref{eq:velcomb}) is tested for the power $P(F,\lambda)=F V(F,\lambda)$ replacing $L$ by $\lambda$, so
\begin{equation}
P(F,\lambda)\simeq b(\lambda)P^{PLS}(F) + [1-b(\lambda)] P^{SR}(F)
\label{eq:power}
\end{equation}
with $b(\lambda)= \lambda^2 (\Lambda/L_l)^2$. 
These results and additional data for $N_f=64$ and $N_f=128$ provided in figure \ref{fig:power_bis} of appendix \ref{sec:ansatz} show the pertinence of eq. (\ref{eq:power}) for a full characterization of the flexibility effects based on analytical expressions, hence relatively straightforward to use in interpreting future in-vitro experimental data.

The long time exponential relaxation of the wall allows to define the concept of chemical friction $\gamma$ 
\begin{equation}
\tau^{OT}(\kappa_T)=-\frac{1}{\kappa_T}\left[\frac{\partial V(F,\kappa_T)}{\partial F}\right]^{-1}_{F=F_s}=\frac{\gamma(\kappa_T)}{\kappa_T}
\label{eq:gamma_chem}
\end{equation}
as the inverse of the slope of the velocity-load relationship at stalling, in principle directly accessible in laboratory experiments. As shown in appendix \ref{sec:tauOT} the $\kappa_T$-dependence can be replaced by the $L_s$-dependence and the chemical friction directly related to the $F$-derivative of $V(F,L)$ at stalling, sensitive to the load-sharing feature of the bundle. The ansatz eq. (\ref{eq:velcomb}) provides
\begin{equation}
\frac{1}{\gamma(L_s)}\simeq b(L_s) \frac{1}{\gamma^{PLS}} + [1-b(L_s)] \frac{1}{\gamma^{SR}}
\label{eq:gamcomb}
\end{equation}
where the $L$-independent friction coefficients of the PLS and the SR Brownian Ratchet models are given by
\begin{eqnarray}
\gamma^{PLS}&=&-\left(\left[\frac{\partial V^{PLS}}{\partial F}\right]_{F=F_s}\right)^{-1}=\frac{N_f k_BT}{d^2 W_0}
\label{eq:gPLS}\\
\gamma^{SR}&=&-\left(\left[\frac{\partial V^{SR}}{\partial F}\right]_{F=F_s}\right)^{-1}= C(N_f,\hat{\rho}_1) \gamma^{PLS}
\label{eq:gBR}
\end{eqnarray}
using the PLS velocity-load relation, eq. (\ref{eq:PLS}), and the SR velocity-load relation in terms of an effective load sharing coefficient $C(N_f,\hat{\rho}_1)$\cite{DCBB.14,PPCR.18}.
For a bundle with $N_f=32$ at $\hat{\rho}_1=2.5$, $\gamma^{PLS}= 32 k_BT/(d^2 W_0)$ and $\gamma^{SR}=281.6  k_BT/(d^2 W_0)$ predicted by eq. (\ref{eq:gBR}) using the approximate $C$ expression in Eq. (C7) of ref. \cite{PPCR.18}. Values of the the chemical friction from our numerical experiments, reported and discussed in appendix \ref{sec:tauOT}, are in semi-quantitative agreement with predictions from eq. (\ref{eq:gamcomb}) based on these values of $\gamma^{PLS}$ and $\gamma^{SR}$. Moreover they confirm   
a progressive decrease of the chemical friction with bundle length which is roughly given by $\gamma \approx \gamma^{PLS}  (L_s/L_l)^{-2}$, for $b(L_s)=(L_s/L_l)^2>0.3$. 

We stress again that probing the bundle relaxation in an optical trap near stalling, for a range of $\kappa_T$ values compatible with the non escaping regime, if possible, would provide an experimental measure of the chemical friction, hence a direct measure  of the $L_s$-dependent load-sharing capacities of a multi-filament actin bundle.

\section{The escaping regime.}
\label{sec:escaping}
In supercritical conditions, the possibility exists for growing flexible filaments hitting a hard obstacle, to undergo a large bending fluctuation and start growing laterally without restraint. 
The onset of such escaping filaments is allowed by our dynamical model. 
Sections \ref{sec:result}-\ref{sec:VFL} were focusing on flexibility effects in the non escaping regime defined by $L<L_l$ defined in eq. (\ref{eq:Ll}). 
In the present section we consider situations where the wall position can sample values $L>L_l$, allowing the system to perform excursions into the escaping regime. 
We classify a filament as escaped (and irremediably lost for the transduction of chemical free energy into mechanical work) when its size gets larger than $z^{*}(L)$ (see its definition and illustration in right panel of figure \ref{fig:sys}).

\begin{figure}[t]
\begin{center}
\includegraphics[width=0.5\columnwidth]{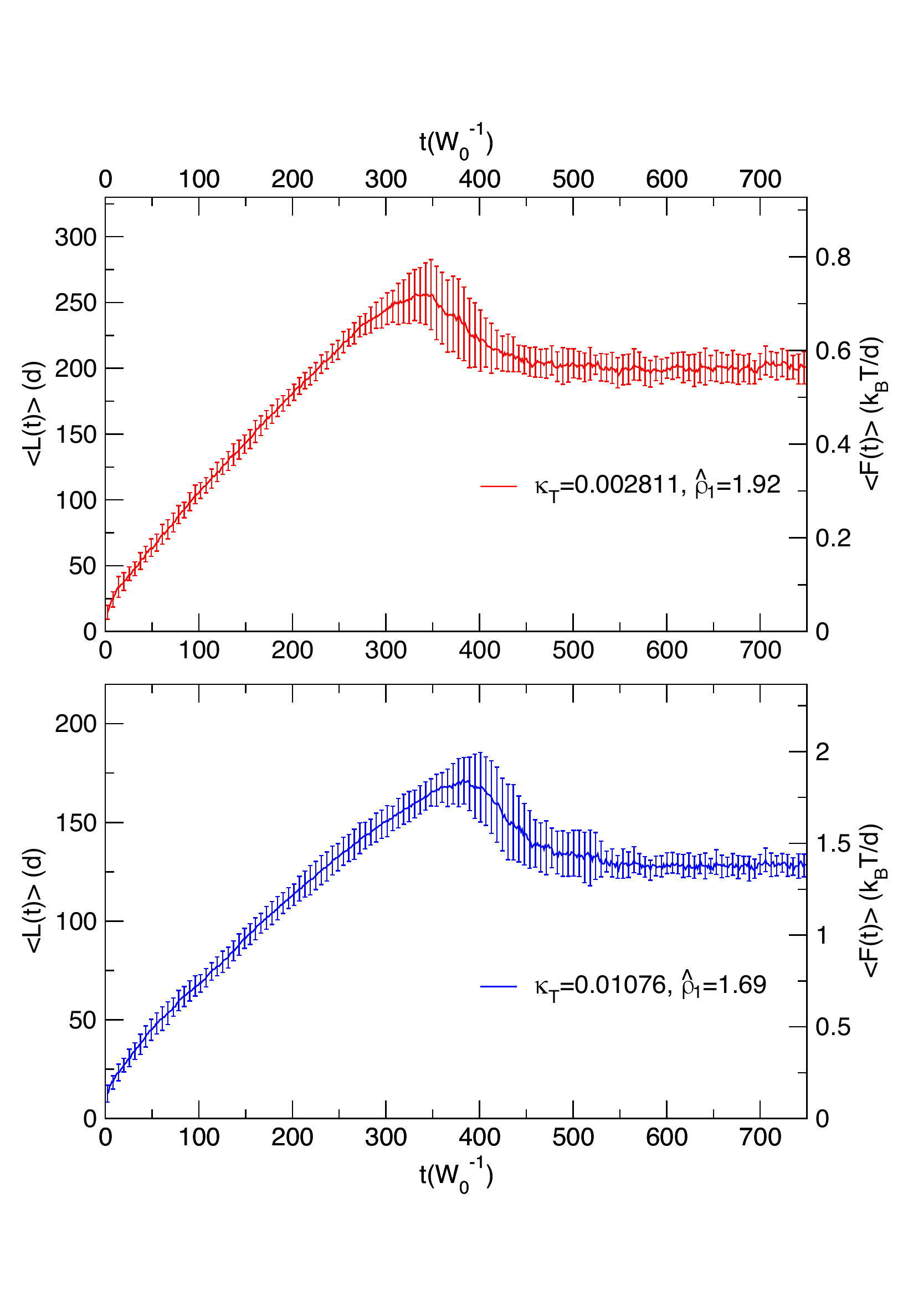}
\caption{Wall relaxation $\langle L(t)\rangle$ for a homogeneous bundle of $N_f=8$ under conditions that mimic the experiments reported in figure 4(a) (upper panel) and 4(b) (lower panel) of ref \cite{Footer.07}, respectively. In both cases an overshooting behavior is observed and corresponds to the occurrence of the escaping filaments. The observed large time plateau values roughly correspond to the mechanical equilibrium between the trap restoring force and the force to bend the eight filaments, while they keep increasing their size unimpeded by the wall.} 
\label{fig:fig5}
\end{center}
\end{figure}

To our knowledge, a unique experimental attempt to follow actin bundle dynamics using the optical trap set-up has been reported to date in ref. \cite{Footer.07}. More specifically, relaxation to stalling for a bundle of 8-10 actin filaments have been monitored. The interpretation of these in-vitro experimental data \cite{Footer.07} turned out to be delicate given the possible bias coming from the presence of escaping filaments. 
It was claimed that in order to probe a genuine polymerization force at stalling the observed final force must be lower than the buckling force of a single grafted actin filament with contour length equal to the trap gap \cite{Footer.07}. 
The relaxation towards a steady value of the trap width for a bundle of 8-10 filaments at $(\hat\rho_1=1.92, \kappa_T=0.0028 k_BT/d^2)$ and at $(\hat\rho_1=1.69, \kappa_T=0.01076 k_BT/d^2)$ was reported in figure 4a and 4b of ref. \cite{Footer.07} respectively. 
Data of the first experiments were discarded while data from the second experiments (and similar cases) were exploited on the basis of the above criteria. 
However, the final value of the trap width measured in the second experiments (figure 4b of ref. \cite{Footer.07}), suggested that the steady value of the force is close to the force needed to stalled a single filament and not a bundle of 8-10 filaments according to eq. (\ref{eq:Fs}). 
This surprising result was interpreted as the result of ``dynamical stalling" where only the longest filament has to be stall as it quickly becomes unstable and depolymerizes rapidly, hence it is replaced by a new longest filament which is stalled in turn by the loaded wall. 
This scenario has not been confirmed so far but it has been conjectured that for many actin filaments to work together and for filaments to avoid buckling, their growing barbed ends at membrane contact must form complexes with processive formin membrane proteins \cite{Blanchoin.14}. 

It is interesting to simulate the experiment within our model and characterize the response, i.e. the time relaxation of the trap. 
We simulated the relaxation of a bundle of 8 filaments at the two experimental conditions described above. 
For each system the dynamics of 32 equivalent replicas, with initial conditions sampled by the equilibrium filament length distribution at $L(0)=5d$, were followed up to the attainment of a steady value for the trap width $L(t)$. 
The results are reported in figure \ref{fig:fig5}. 

In both cases we observe a non-monotonous behavior of $\langle L(t)\rangle$ at variance with the results reported in figure \ref{fig:fig4}. 
The observed overshoot is, in both experiments, the signature of the occurrence of escaping filaments. 
Indeed, all filaments of all replica became escaping in the time interval between the end of the first, almost linear, growth and the beginning of the plateau observed at larger times. 
In this time interval we observe an increase of the variance signaling the random character of the filament escaping times in different replica. 
The asymptotic value of the trap width corresponds to the load exerted by the trap to equilibrate the elastic force provided by eight very long d-WLC bent at 90 degrees within the width of the trap. 
Equation (\ref{eq:Lesc}) predicts $L_{esc}\simeq 230d$ and $L_{esc}\simeq 147d$ for the two experiments respectively, in qualitative agreement with the observed values. 
The corresponding threshold values for the trap strength from eq. (\ref{eq:kappa_esc}) are $\bar{\kappa}_T=0.064k_BT/d^2$ and $\bar{\kappa}_T=0.046k_BT/d^2$ both rather larger than the employed values. 
According to eq.(\ref{eq:Ll}), the crossover distance to the escaping regime in the two cases is $L_l(1.92)\simeq 91d$ and $L_l(1.69)\simeq 101d$ rather shorter than the observed $L_{esc}$. 
From the analysis of the trajectories, in the upper panel of figure \ref{fig:fig5} ($\kappa_T=0.002811 k_BT/d^2$) we detect the first escaping filaments after an average time $t_1=381 W_0^{-1}$ ($\sigma_{t_1}=23 W_0^{-1}$) while all $N_f=8$ filaments are found to escape after an average time $t_2=413.0 W_0^{-1}$ ($\sigma_{t_2}=23 W_0^{-1}$). 
In the lower panel ($\kappa_T=0.01076 k_BT/d^2$), we have corresponding times $t_1=437 W_0^{-1}$ ($\sigma_{t_1}=28 W_0^{-1}$) and $t_2=498 W_0^{-1}$ ($\sigma_{t_2}=51 W_0^{-1}$). 
The onset of escaping filaments thus takes place essentially during the backward displacement of the wall, just after the overshoot at $L$ values much higher than $L_{l}$. 
This is probably due to a kinetic effect linked to the initial fast motion of the wall, $V \approx d W_0$ (higher panel) and $V \approx 0.5~dW_0$ (lower panel), close to the free growth velocity at zero load $(U_0-W_0)d$ (which is $0.92~dW_0$ and $0.69~dW_0$ respectively). 
Such large wall velocity probably prevents filaments to enter earlier in the escaping regime $i>z^{*}(L)$.

In ref. \cite{PPCR.16} we have already pointed out that for a bundle of $8$ filaments growing at $(\hat\rho_1=1.69, \kappa_T=0.01076 k_BT/d^2)$ (second experimental case) of the expected stalling force from eq. (\ref{eq:Fs}) would imply an average trap gap much larger than the crossover $L_l$ and hence the relaxation process could not avoid incurring in the escaping regime. 
Our new results confirm the occurrence of dynamical instabilities when filaments become escaping, as the trap moves under the combined action of the filaments force and the external load. 

For the experimental data at $\kappa_T d^2/k_BT=0.01076$ (figure 4b in \cite{Footer.07}), the plateau is reached asymptotically after a growth characterized by a relaxation time $\tau^{OT}\simeq 55s$ (disregarding that the growth is affected by unexplained oscillations). 
Hence, the experimental data are compatible with the relaxation of a single filament, not only for the plateau value but also for the longest relaxation time as it is close to $\tau^{OT}= k_BT(W_0 \kappa_T d^2)^{-1}=66 s$ predicted by using the experimental $\kappa_T$ value, $W_0=1.4s^{-1}$ and eqs. (\ref{eq:tau_L})-(\ref{eq:gPLS}) with $N_f=1$.\footnote{Note that for a single filament, the PLS and the SR models coincide.}

We want to emphasize that our present results concern a model based on a single species of actin monomer complexes (no hydrolysis effects) with a depolymerizing rate supposed to be unaffected by the presence of the compressional force due to the obstacle. 
Therefore it does not contain the ``dynamical stalling" phenomenon invoked to explain the puzzling experimental results. It would be interesting to perform a similar study with a model where a compression dependent depolymerization rate is considered. However, at present, the experimental input about this dependence is missing.

\section{Discussion and conclusion}
\label{sec:concl}

\begin{figure}[t]
\begin{center}
\includegraphics[width=0.6\columnwidth]{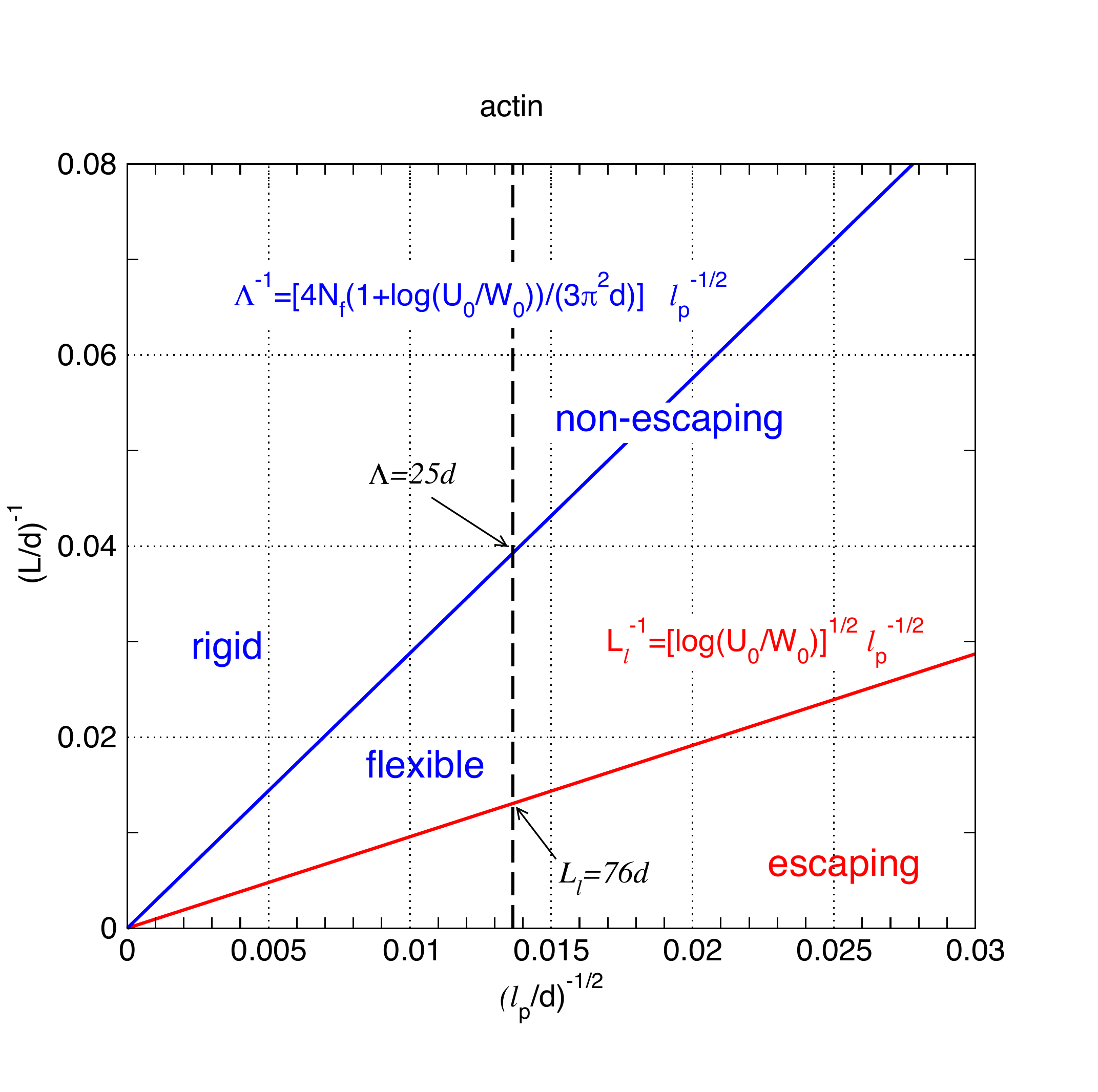}
\caption{The three different regimes for a bundle of filaments with finite flexibility indicated by the filament persistence length, illustrated here for $N_f=32$ and $\hat{\rho}_1=2.5$. The case of actin $\ell_p=5370d$ is indicated by a vertical dashed line. The non escaping regime for a homogeneous bundle is further divided into a rigid filament regime and a flexible filament regime by the line at $L=\Lambda$ given in the text. The escaping regime is met for $L=L_l$ with $L_l=\left(\ell_pd /\ln{\hat{\rho}_1}\right)^{1/2}$.} 
\label{fig:fig3}
\end{center}
\end{figure}

Using a realistic model of semi-flexible filaments, we have introduced, in a rather general fashion, the effects of flexibility on the dynamics of actin bundles in supercritical conditions pressing against a loaded obstacle. 
Applying an external load that mimics the optical trap apparatus in realistic conditions we have followed the non-equilibrium relaxation of the trap under the combined action of the polymerization force of a staggered actin bundle and the external load. 
The relaxation of the trap is found to be mostly adiabatic as the reorganization of the filament sizes against the loaded wall is one-two orders of magnitude faster than the time of the load. This fundamental property allows to consider most states visited during the dynamical relaxations as non-equilibrium stationary states for the bundle and, by eliminating the time from the analysis of the results, to generalize the well known velocity-load relationship of rigid bundles to include flexibility.
The velocity $V(F,L)$ now depends in addition on the bundle length $L$, which greatly enriches the theoretical scenario. 
The new fundamental features are: 
i) a new characteristic length $\Lambda \approx (d \ell_p/N_f)^{1/2}$ (see eq. (\ref{eq:Lstar})), a bundle property that marks the transition between SR behavior and flexible bundle behavior, so the system properties are more adequately described in terms of the reduced length $\lambda=L/\Lambda$; 
ii) the enhancement of the power of transduction of chemical energy into mechanical work for $\lambda>1$; 
iii) the occurrence of the escaping regime beyond some threshold distance $L_l$ related to single filaments characteristics.
Figure \ref{fig:fig3} illustrates these features and their dependence on the filament persistence length $\ell_p$. 
We show the three regimes: 1) a rigid regime at $L<\Lambda$ ($\lambda<1$) where the flexibility effect is negligible; 2) the intermediate regime where flexibility effects are present with an increase of the obstacle velocity with flexibility (i.e. bundle length at fixed $\ell_p$) at fixed external load; 3) the escaping regime at large enough trap width ($L>L_l$). We see that the non escaping flexible regime gets wider for increasingly flexible filaments.
\footnote{For an unstaggered bundle, flexibility would add the same new polymerization mechanism when a compression of d would become of the order of $k_BT$. But now $\Lambda \sim L_l$ and the new non escaping flexible regime is absent.}  
Let us consider the ratio $L_l/\Lambda=\left[ 4/3\pi^2~N_f~(1+\ln\hat\rho_1)/\ln\hat\rho_1\right]^{1/2}$, which is independent of $\ell_p$. 
In supercritical conditions ($\hat\rho_1>1$) $L_l/\Lambda\geq \left(4/3\pi^2\right)^{1/2} N_f^{1/2}$, therefore a wide flexible regime before escaping requires bundles with a large number of filaments. 

Using the optical trap set up for different values of $\kappa_T$ we characterized the three regimes and established their boundaries. 
We have shown that the power of the bundle is strongly increased by the flexibility in the intermediate regime and can almost cover the entire gap between the rigid bundle behavior (SR) and the mean field \textit{Perfect Load Sharing} (PLS) behavior, which appears to be an upper bound for a many flexible filaments bundle as the filaments in the bundle approaches the escaping regime. 
The spectacular power increase with $L$ at a given load $F$
is the result of an improved work sharing capacity of the bundle due to the increasing fraction of filaments pressing on the wall ($\propto L^2$) when developing the polymerization force opposing the load $F$. 
A simple \textit{ansatz} combining linearly the PLS and SR behaviors provides a satisfactory description of the data.   
Our work shows how filament flexibility could be considered in interpreting future experiments, in a way which enriches considerably the present dominant  theoretical model \cite{Mogilner.Rubinstein}. 
Despite a consistent treatment of filament flexibility including the escaping regime, our model is unable to reproduce the in-vitro experimental results of ref. \cite{Footer.07} which remain unexplained. 

\section*{Acknowledgments}
C.P. and J.P.R. are grateful to J. Baudry, J.F. Joanny et D. Lacoste for useful discussions. We thank G. Destr\'ee and P. Pirotte for technical help. J.P.R. thanks the financial support and hospitality of the University of L'Aquila during a three months visit. C.P. is supported by the Agence Nationale de la Recherche (ANR), France under the project ``HyLightExtreme''. Computational resources have been partially provided by the ``Consortium des \`Equipements de Calcul Intensif (C\`ECI)", funded by the Fonds de la Recherche Scientifique de Belgique (F.R.S.-FNRS) under Grant No. 2.5020.11.

\begin{appendix}

\section{Tables of symbols and physical properties }\label{app_table}
In this appendix we collect, in three tables, all symbols used in the present text. Table \ref{tab1} is devoted to macroscopic quantities, table \ref{tab2} to microscopic quantities and table \ref{tab3} to time scales.
\begin{table}[h]
\caption{Macroscopic/mesoscopic quantities}
\renewcommand*{\arraystretch}{1.8}
\begin{center}
\begin{tabular}{|>{\centering\arraybackslash}p{4cm}|>{\centering\arraybackslash}p{11cm}|}
       \hline
       \hline
       \textbf{Symbol}     & \textbf{Meaning} \\
       \hline
       $\rho_1$   & Bulk free actin monomer density.  \\
	   $N_f$      & Number of filaments in the bundle. \\
	   $U_0=k_{on} \rho_1$ & Bulk polymerization rate per filament tip. \\
       $W_0=k_{off}$ & Bulk depolymerization rate per filament tip. \\
       $\rho_{1c}=\dfrac{k_{off}}{k_{on}}$ & Critical free monomer density. \\
       $\hat{\rho}_1=\dfrac{\rho_1}{\rho_{1c}}=\dfrac{U_0}{W_0}$ & Reduced density. \\
       $L$        & Bundle length and/or optical trap displacement with respect to the grafting wall position. eq. (\ref{eq:Ll}) and caption of figure \ref{fig:sys}. \\
       $\kappa_T$ & Trap strength. Caption of figure \ref{fig:sys}.\\
       $F=\kappa_T L$   & Load force in the optical trap. Caption of figure \ref{fig:sys}.\\
       $V=\dfrac{dL}{dt}$    & Obstacle velocity. \\
       $F_s$, $L_s=\dfrac{F_s}{\kappa_T}$  & Stalling force and stalling length. eq. (\ref{eq:Fs}).\\
       $L_l$ & Crossover bundle length between non escaping and escaping regimes. eq. (\ref{eq:Ll}).\\
       $\xi$ & Friction coefficient of the wall in the solution. eq. (\ref{eq:Langevin}).\\
       $\gamma=-\left[\dfrac{\partial V(F)}{\partial F}\right]_{F_s}^{-1}$ & Effective chemical friction coefficient near stalling. eq. (\ref{eq:gamma_chem}).\\
       $P=V F$ & Transduction power of the bundle. \\
       $\bar{L}(n)$ & Filament length at which a polymerization involving a wall compression of $nd/N_f$ becomes thermally activated. eq. (\ref{eq:lbar}).\\
       $\Lambda$ & Bundle size threshold between the rigid bundle regime and the flexible (non escaping) regime. eq. (\ref{eq:Lstar}).\\
       $\lambda=\dfrac{L}{\Lambda}$ & Flexibility parameter: rigid regime $\lambda<1$, flexible regime $\lambda>1$. Caption of figure \ref{fig:fig1}.\\
       \hline
       \hline
\end{tabular}
\end{center}
\label{tab1}
\end{table}%

\begin{table}[h]
\caption{Microscopic parameters and intermediate quantities}
\renewcommand*{\arraystretch}{1.6}
\begin{center}
\begin{tabular}{|>{\centering\arraybackslash}p{4cm}|>{\centering\arraybackslash}p{11cm}|}
       \hline
       \hline
       \textbf{Symbol}     & \textbf{Meaning} \\
       \hline
       $d=2.7nm$ & F-actin incremental length per additional monomer in the double strand actin filament. \\
       $\ell_p=14.5 \mu m$ & F-actin persistence length. \\
       $h_n$ & Seed longitudinal position of filament $n$ with respect to grafting plane ($-d/2 \leq h_n \leq d/2$). eq.(\ref{eq:seeds}).\\
       $j_n$ & Number of monomers in filament $n$. Text below eq.(\ref{eq:seeds}).\\
       $L_{cn}$ & $n$-th filament contour length $L_{cn}=(j_n-1)d$. Text below eq.(\ref{eq:fbun}). \\
       $L_n$ & $n$-th filament seed-wall distance $L_n=L-h_n$. Text below eq.(\ref{eq:fbun}).\\
       $z(L,h_n)$ & Largest monomer size of filament $n$ for which bending fluctuations do not involve any direct interactions with the wall. Caption of figure \ref{fig:sys}. \\
       $z^*(L,h_n)$ & Assumed cross over filament size between the non-escaping ($j_n \leq z^*$) and the escaping ($j_n > z^*$) regime. Caption of figure \ref{fig:sys}. \\
       $f_b=\dfrac{\pi^2}{4} \frac{k_BT \ell_p}{\tilde{L}^2}$ & Buckling force for a filament of length $\tilde{L}$. eq.(\ref{eq:buckling}). \\
       $-k_BT \ln{\alpha_j(L)}$ & Potential of mean compression force for a particular filament of fixed size $j$ and seed at $h=0$ facing a wall at $L$. Eq.(\ref{eq:fe}) for $\alpha$ definition.\\
       $w_j$ & Total free energy (chemical+compressional) of a particular filament of size $j$. Eq.(\ref{eq:fe}) and figure \ref{fig:fig6}.\\ 
       $\bar{f}_j$ & Force deriving from $w_j$. Eq.(\ref{eq:force_i}).\\
       $F_{bun}$ & Microscopic bundle force. Eq.(\ref{eq:fbun})\\
       \hline
       \hline
\end{tabular}
\end{center}
\label{tab2}
\end{table}%

\begin{table}[h]
\caption{Time scales}
\renewcommand*{\arraystretch}{1.6}
\begin{center}
\begin{tabular}{|>{\centering\arraybackslash}p{4cm}|>{\centering\arraybackslash}p{11cm}|}
       \hline
       \hline
       \textbf{Symbol}     & \textbf{Meaning} \\
       \hline
       $\tau_{chem}=W_0^{-1}$ & Chemical characteristic time. Discussion justifying eqs.(\ref{eq:fbun},\ref{eq:fe},\ref{eq:force_i}).\\ 
       $\tau_D=\dfrac{\xi d^2}{k_BT}$ & Local wall diffusion characteristic time. Discussion justifying eqs.(\ref{eq:fbun},\ref{eq:fe},\ref{eq:force_i}).\\
       $\tau^{OT}$ & Longest relaxation time of a bundle in supercritical conditions subject to an optical trap compression load, related to the relaxation of the bundle in optical trap towards equilibrium ($\kappa_T L_s=F_s$). Eqs.(\ref{eq:Lt}), (\ref{eq:tauOT}).\\
       \hline
       \hline
\end{tabular}
\end{center}
\label{tab3}
\end{table}%

\section{Further support to the validity of the ansatz eq.(\ref{eq:power}) to interpret the power of a staggered flexible bundle.}
\label{sec:ansatz}

\begin{figure}[h]
\begin{center}
\includegraphics[width=0.6\columnwidth]{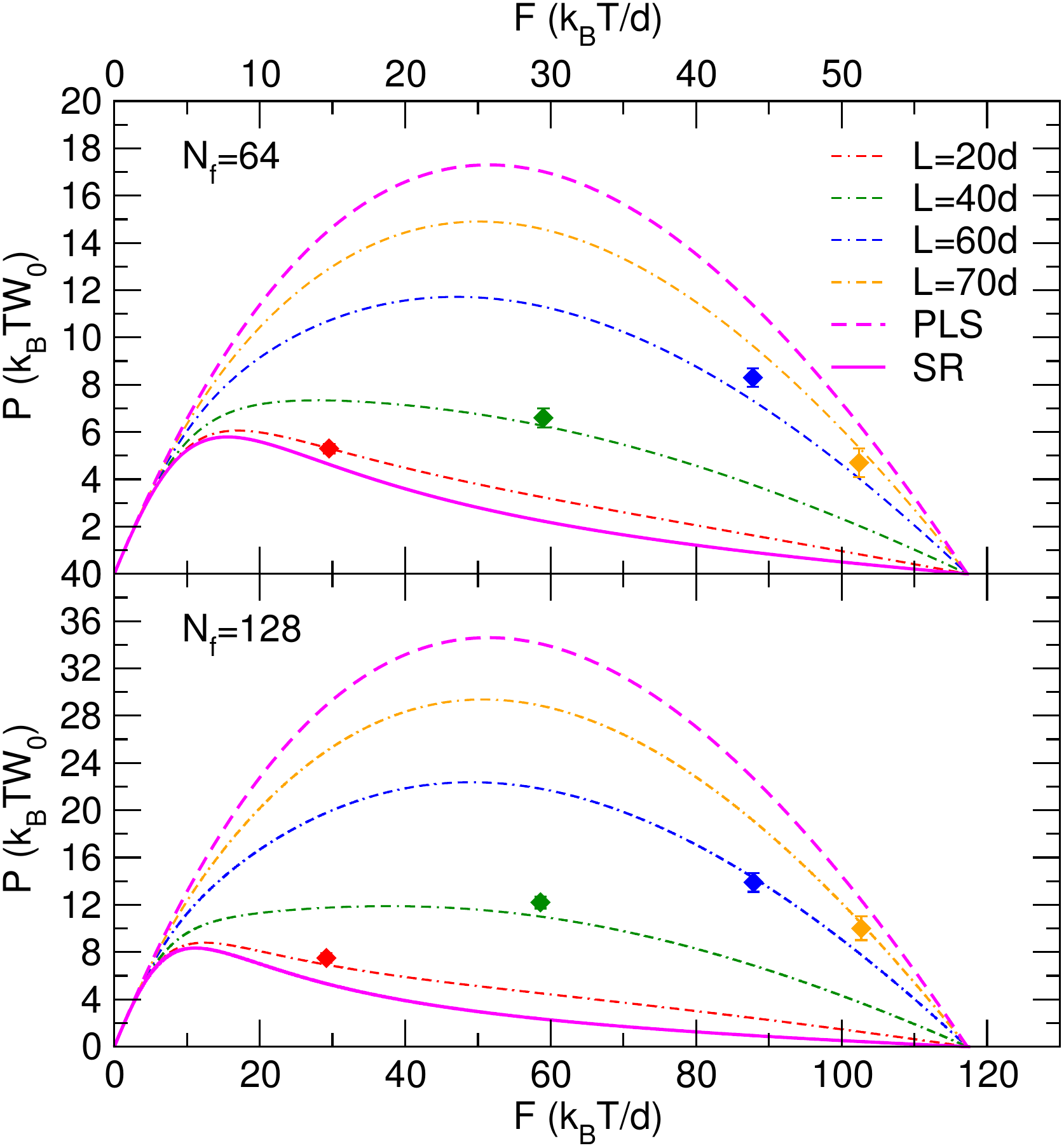}
\caption{Flexible bundles relaxation experiments against an harmonic load with trap strength $\kappa_T=0.011453 N_f$ chosen to cover the same $L$ window between initial size $L_0\approx 5d$ up to a common equilibrium value $L^{eq}\approx 80 d$. The data points correspond to successive situations for increasing bundle size values during the fixed $\kappa_T$ relaxations for $N_f=64$ (upper panel) and $N_f=128$ (lower panel). Each point represents $P=F V$  versus F where $V$ is the instantaneous velocity and $F=\kappa_T L$ is the instantaneous external force applied to the bundle by the loaded wall. The plot show the power predicted by eq.(\ref{eq:power}). }
\label{fig:power_bis}
\end{center}
\end{figure}

In this appendix we report results for bundles with different number of filaments at the same reduced trap strength, to further support our analysis of the velocity-load-length relationship, in particular the weighted average between the SR and the PLS behaviors given by the ansatz eq.(\ref{eq:power}). In figure \ref{fig:power_bis}, we compare data for the optical trap relaxation of bundles of $64$ and $128$ flexible filaments to the prediction eq.(\ref{eq:power}) to stress its general validity, beyond the single case $N_f=32$ illustrated in the main text.

\section{Relaxation time of a staggered flexible bundle in optical trap and chemical friction.}
\label{sec:tauOT}

In this appendix we show how to formally go from the $\kappa_T$-dependent chemical friction defined in eq.(\ref{eq:gamma_chem}) to the $L$-dependent chemical friction. 
Applying the variable change $y=g(x,z)$ to a function of two variables $f(x,y)$, one has the general relationship $\left[\frac{\partial f}{\partial x}\right]_{z}=\left(\left[\frac{\partial f}{\partial x}\right]_{y} + \left[\frac{\partial f}{\partial y}\right]_{x} \left[\frac{\partial g}{\partial x}\right]_{z}\right)$. For the specific $g(x,z)=x/z$, identifying $x=F$, $y=\kappa_T$, $z=L$, one gets
\begin{equation}
\left[\frac{\partial V}{\partial F}\right]_{\kappa_T}=\left[\frac{\partial V}{\partial F}\right]_{L}+ \frac{1}{L}\left[\frac{\partial V}{\partial \kappa_T}\right]_{F}
\label{eq:thermo1}
\end{equation}
The second term in the rhs of eq. (\ref{eq:thermo1}), which is strictly zero in the rigid filament case, is negligible for a flexible bundle close to stalling as it is linked to the very small change of $F_s$ with equilibrium bundle length fixed by $\kappa_T$, and we reduce to
\begin{equation}
\left[\left[\frac{\partial V}{\partial F}\right]_{\kappa_T}\right]_{F=F_s}\approx \left[ \left[\frac{\partial V}{\partial F}\right]_{L}\right]_{F=F_s}
\label{eq:thermoZ}
\end{equation}
Hence we can replace $\kappa_T$ with $L_s$ in eq. (\ref{eq:gamma_chem}) to define 
\begin{equation}
\tau^{OT}(L_s)=-\frac{1}{\kappa_T}\left[\frac{\partial V(F,L)}{\partial F}\right]^{-1}_{F=F_s}=\frac{\gamma(L_s)}{\kappa_T}
\label{eq:tau_L}
\end{equation}
The chemical friction coefficients $\gamma(\kappa_T)$ from eq. (\ref{eq:gamma_chem}) and $\gamma(L_s)$ from eq. (\ref{eq:tau_L}) are equal when referring to a specific equilibrium state linking the equilibrium wall position $L_s$ and the trap strength $\kappa_T$, at a given ($N_f,\rho_1$) state point. 
\begin{figure}
\begin{center}
\includegraphics[width=0.6\columnwidth]{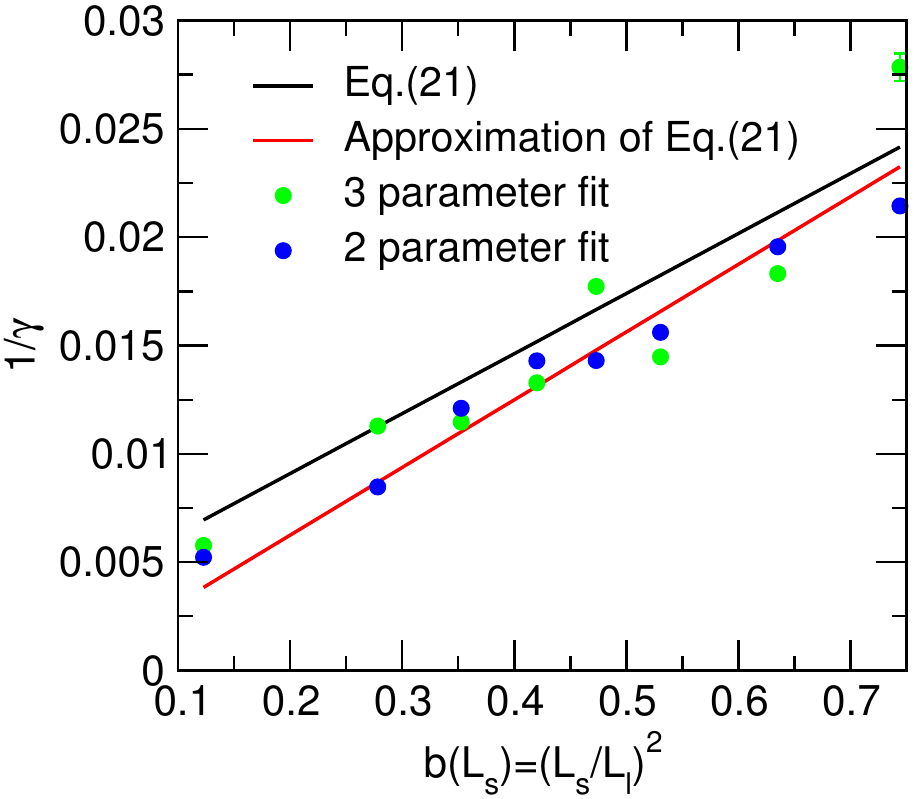}
\caption{Inverse of chemical friction ($1/\gamma$) versus $b(L_s)=(L_s/L_l)^2$ from the long-time relaxation of a bundle of $32$ flexible actin filaments in optical trap relaxation data in the non escaping regime, illustrated in figures \ref{fig:fig2}-\ref{fig:fig4}. 
The two parameters fit ($p_1,p_2$) consists in extracting an effective relaxation time $\tau^{OT}=p_2$ using the form $\langle L(t)\rangle=L_s-p_1 \exp{(-t/p_2)}$ where $L_s$ is the exact theoretical prediction of the equilibrium position of the trap \cite{PPCR.16}, in an adjusted time window where $T_{min}$ is selected empirically to get an optimal single exponential fit for the long time behaviour. 
The three parameters fit ($p_1,p_2,p_3$) consists in extracting an effective relaxation time $\tau^{OT}=p_2$ using the best fit with the form $\langle L(t)\rangle=p_3-p_1 \exp{(-t/p_2)}$ in a fixed time window $[T_{min},T_{max}]=[400,600]W_0^{-1}$. 
The continuous curves are the predictions of eq.(\ref{eq:gamcomb}) and its approximation $\gamma=(L_l/L_s)^2 \gamma^{PLS}$ (see text)}.
\label{fig:friction}
\end{center}
\end{figure}

We analyzed the flexibility influence on the chemical friction as a function of the bundle equilibrium length $L_s$.
Data for a bundle with $N_f=32$ at various values of $L_s$ in the non escaping regime ($0.5<\kappa_T <1.1$) from our SD simulations are reported in figure \ref{fig:friction} in terms of the weight $b=(L_s/L_l)^2$ of the PLS contribution. The weighted average ansatz of eq. (\ref{eq:gamcomb}) together with the approximation $\gamma=(L_l/L_s)^2 \gamma^{PLS}$ derived from eq.(\ref{eq:gamcomb}) by neglecting $(1/\gamma^{SR})$ versus $(1/\gamma^{PLS})$, are indicated to illustrate the $\gamma \approx 1/L_s^2$ behavior. 
Extracting the relaxation time directly from the numerical relaxations is not trivial since it requires small statistical errors in the long time region where the deviations from the asymptotic trap length become vanishingly small. With the typical noise level of our data the value of the relaxation time depends substantially on the time window of the fit. 
Two estimates of the chemical friction extracted from SD results are shown as illustrated in the caption of figure \ref{fig:friction}. The results confirm a progressive decrease of the friction as the bundle length increases, hence an increase of the load-sharing capacity of the bundle.

\section{Experimental time scales}
\label{sec:char_time}

The overdamped limit of the Langevin equation eq.(\ref{eq:Langevin}) is justified by the very fast inertial relaxation time $\tau^{inertia}=M/\xi=2 \times 10^{-7}$s estimated according to the experimental setup \cite{Footer.07} for a micron radius bead $M=\frac{4 \pi}{3} R^3 \rho \approx 4.2\times 10^{-15}$Kg and from Stokes law $\xi=1.9 \times 10^{-8}$ Js/m$^2$. Other relevant time scales are much longer: the trap relaxation time in absence of bundle is $\tau_{free}=\xi/\kappa_T=2.5$ ms (using a typical value $\kappa_{T}=0.008$ pN/nm \cite{Footer.07}) and the chemical reactions time scale is $\tau_{chem}=W_0^{-1}\approx 0.7$~s \cite{b.Howard}.

Another relevant time scale which characterize the dynamics is the diffusive one related to the diffusive motion of the wall over the length corresponding to the size of the (de)polymerization unit and estimated by $\tau_D=d^2/D$ where $D$ is the wall diffusion coefficient. 
As in reference \cite{PPCR.18}, let's define  $\epsilon=\tau_{chem}/\tau_D$. 
Experimentally, $\epsilon=4.7\times10^{-5}$ for actin, using $\xi=1.9\times10^{-8}$ Js/m$^2$, $W_0=1.4$ s$^{-1}$, $d=2.7$ nm and $k_BT=4.14\times10^{-21}$J. 
Such a small value cannot be adopted in the simulations for computational inefficiency reasons. 
Computer time would be essentially spent to study the wall relaxation next to a bundle with quasi-fixed filament sizes while our main interest is the force-velocity requiring both wall and filament size sampling. 
Hence, as in our recent investigation of the rigid bundle case \cite{PPCR.18}, we have adopted the value $\epsilon=4.7\times10^{-2}$, small enough to be representative of the experimental wall dynamics with $\epsilon\simeq5\times10^{-5}$.

The longest intramolecular relaxation time of a grafted WLC with contour length $L_c$ is\cite{TPFF08}
\begin{equation}
\tau_{intra}=\left(\frac{2}{\pi}\right)^4\frac{\xi' L_c^4}{k_BT \ell_p}
\label{tau_intra}
\end{equation}
where $\xi'$ is a drag coefficient per unit length. 
$\xi'$ can be estimated by \cite{TPFF08} 
\begin{equation}
\xi'_0~=~\frac{4 \pi \nu}{(\ln{(L_c/a)}+2 \ln{2} -0.5)}
\end{equation}
where $\nu$ is the solvent viscosity and $a$ is the diameter of the rod modeling F-actin hydrodynamically ($a=5.4~$nm$=2d$ \cite{GKFBS96}).
Writing $L_c=(j-1)d$ for the filament contour length with $j$ articulation points, one gets 
\begin{eqnarray}
\tau_{intra}&\simeq&\frac{64 (j-1)^4}{(\ell_p/d) \pi^3 [\ln\frac{j-1}2+2\ln2-0.5]}\frac{\nu d^3}{k_BT} \nonumber\\
&=&\left(\frac{(j-1)^4}{\ln\frac{j-1}2+2\ln2-0.5}\right)\cdot1.83\times10^{-12}\text{ s}\qquad
\end{eqnarray}
using $\nu=0.001$, $d=2.7\times10^{-9}$, $k_BT=4.14\times10^{-21}$ in MKS units. Thus one finally gets estimates $\tau_{intra}=3\times10^{-6}~$s for $j=51$ and $\tau_{intra}=4\times10^{-5}~$s for $j~=~101$, hence showing that the ratio $\tau_{intra}/\tau_{chem}~\leq~5\times10^{-5}$ as long as $j~\leq~100$.

\section{Explicit wall stochastic dynamics Algorithms (EWA)}
\label{sec:EWA}

To get a numerical solution of the model described above using a Langevin equation approach, we adopt an Explicit Wall stochastic dynamics Algorithm (EWA) developed in ref. \cite{PPCR.18} to follow the dynamics of a bundle of rigid filaments in optical trap, that we adapt below to the flexible case.
We start from the equivalent Fokker-Planck formulation of the time evolution of the joint distribution function $P_{j_1,j_2,....j_{N_f}}(L,t)$. 
Discretizing the wall position variable, the dynamics reduces to a multi-variables continuous-time Markov chain for which standard algorithms are available to generate representative stochastic trajectories.   
 
The Fokker Planck equation describing the evolution of the joint probability distribution of the variables $(L,\{j_n\}_{n=1,N_f})$ in the case of flexible filaments reads:
\begin{eqnarray}
&&\frac{\partial P_{j_1,\dots,j_{N_f}}(L,t)}{\partial t} =-\frac{\partial}{\partial L} J_{j_1,\dots,j_{N_f}}(L,t)\nonumber\\
 &+&\sum_{n=1}^{N_f}U_n(j_n-1,L)(1-\delta_{2,j_n})P_{j_1,\dots,j_n-1,\dots,j_{N_f}}(L,t)\nonumber\\
 &-&\sum_{n=1}^{N_f}U_n(j_n,L)P_{j_1,\dots,j_n,\dots,j_{N_f}}(L,t)\nonumber\\
 &+&\sum_{n=1}^{N_f}W_n(j_n+1,L)P_{j_1,\dots,j_n+1,\dots,j_{N_f}}(L,t)\nonumber\\
 &-&\sum_{n=1}^{N_f}W_n(j_n,L)P_{j_1,\dots,j_n,\dots,j_{N_f}}(L,t)
\label{eq:FPeqF}
\end{eqnarray}
where $U_n$ and $W_n$ are the (de)polymerization rates given in eqs.(\ref{eq:U}),(\ref{eq:W}), and the probability current density is
\begin{eqnarray}
&&J_{j_1,\dots,j_{N_f}}(L,t)=-D\frac{\partial P_{j_1,\dots,j_{N_f}}(L,t)}{\partial L}\nonumber\\ &-&\frac{D}{k_BT}\left(\sum_{n=1}^{N_f}\frac{\partial w_{j_n}(L_n,\hat\rho_1)}{\partial L}+ \kappa_TL\right)P_{j_1,\dots,j_{N_f}}(L,t).\nonumber\\
\label{eq:pcdF}
\end{eqnarray}
The boundary conditions for these equations are different from the rigid case, where the most advanced tip's position results in a reflecting boundary condition for the wall motion: here, conversely, filaments can bend and let the wall move to positions $L<(j_n-1)d$. 
Two kinds of boundary conditions can be chosen: 
\begin{itemize}
\item [a)] filament sizes are limited to $z^{*}(L)=\text{int}((\pi L)/(2d))$, and hence Gholami's expression for the free energy $w_{j_n}(L_n,\hat\rho_1)$ \cite{Frey.06}, which is valid in the weak bending regime, can be used. This leads to the stationary state described in reference \cite{PPCR.16} by equilibrium statistical mechanics.
\item [b)] filament sizes are not limited, and escaping filaments can appear. In this case, depending on $\kappa_T$, the evolution of the system can lead to no true stationary state, with a finite average wall position (see figure \ref{fig:fig5} and appendix \ref{sec:WLC}) and filament sizes keeping on growing unhindered. 
\end{itemize}
These latter \textit{open} boundary conditions (no real stationary state) have been used in this paper. We used for the mean force potential $w_{j_n}(L_n,\hat\rho_1)$ an expression which is valid also beyond the weak bending regime, as detailed in appendix \ref{sec:WLC}.

To solve the set of Fokker-Planck eqs. (\ref{eq:FPeqF}),(\ref{eq:pcdF}) numerically, we have discretized the wall position, allowing it to move by forward or backward jumps on a grid of step $\delta'=d/M$, with $M\gg1$ integer. 
The SD trajectory follows the time evolution of the ensemble of $(N_f+1)$ integer variables: the filament sizes, $j_n(t)$ $(n=1,N_f)$, and the index $k(t)$, which gives the wall position $L(t)=k(t) \delta'$. 
These variable can change by $\pm 1$ with elementary rates to be specified: 
as for the filaments, the (de)polymerization rates are given by eqs.(\ref{eq:U}),(\ref{eq:W}), while the rates for the stochastic jumps of the wall position are obtained through the discretization of the diffusive part of eq. (\ref{eq:FPeqF})\cite{PPCR.18}, and read 
\begin{align}
W_{+1}& = \frac{D}{\delta'^2}\frac{\beta \Delta U_{+1}}{[\exp{(\beta \Delta U_{+1})}-1]}\nonumber\\
W_{-1}& = \frac{D}{\delta'^2}\frac{\beta \Delta U_{-1}}{[\exp{(\beta \Delta U_{-1})}-1]}
\label{eq:W+-}
\end{align}
where the global potential energy $U$ incorporates the optical trap term and the mean force potential eq.(\ref{eq:fe}) of all passive filaments in terms of an index $k_n$ defined as $L_n=k\delta'-h_n$. 
We have
\begin{align}
\beta \Delta U_{+1}&=\sum_{n=1}^{N_f}\left[-\ln{\left(\frac{\alpha_{j_n}(L_n+\delta')}{\alpha_{j_n}(L_n)}\right)}\right]+ \frac{\beta\kappa_T}{2}  [(L+\delta')^2-L^2] \\
\beta \Delta U_{-1}&=\sum_{n=1}^{N_f}\left[-\ln{\left(\frac{\alpha_{j_n}(L_n-\delta')}{\alpha_{j_n}(L_n)}\right)}\right]+ \frac{\beta\kappa_T}{2}  [(L-\delta')^2-L^2]
\label{eq:pot_ch}
\end{align}
Given the state of the $N_f+1$ discrete variables $\left(\{j_n(t_m)\}_{n=1,\dots,N_f},k(t_m)\right)$ after $m$ steps, the next state after $m+1$ steps is given by changing one of these variables by $\pm$ one unit, according to stochastic rules based on all the above elementary rates \cite{GG.07}. 
Random sampling fixes both the variable that is going to change and the associated time of occurrence $t_{m+1}-t_m$ as follows.
Let's denote by $\{T_i\}_{i=1,\dots,2(N_f+1)}$ the ordered rates of the $2(N_f+1)$ possible single variable changes:
\begin{itemize}[wide,labelindent=0pt,leftmargin=0.5in]
\item[\emph{depolymerizations}:] $T_{2n}=W_n(j_n(m),k)$ using eq.(\ref{eq:W}) 
\item[\emph{polymerizations}:] $T_{2n-1}=U_n(j_n(t_m),k(t_m))$ using eq.(\ref{eq:U})
\item[\emph{wall jumps}:] $T_{2N_f+1}=W_{+1}(\{j_n(t_m)\},k(t_m))$ and $T_{2N_f+2}=W_{-1}(\{j_n(t_m)\},k(t_m))$ using eq.(\ref{eq:W+-})
\end{itemize}
and let's define the vector $\{S_\ell=\sum_{i=1}^\ell T_{i}\}_{\ell=1,2(N_f+1)}$.
Two independent random numbers $(r_1,r_2)$ are sampled from a uniform distribution between $0$ and $1$ to get the time to the next move $\tau=t_{m+1}-t_m$ and the nature of the move, defined by the index $\ell$ among the ordered rate vector $T$ ($1 \leq\ell\leq 2(N_f+1)$):
\begin{align}
\tau&=\frac{1}{S_{2(N_f+1)}} \ln{(1/r_1)}\label{eq:r1}\\
\frac{S_{\ell-1}}{S_{2(N_f+1)}} &< r_2 \leq \frac{S_{\ell}}{S_{2(N_f+1)}}.
\label{eq:r2}
\end{align}
The set of $N_f+1$ microscopic variables are stored at equally spaced times for analysis.

\section{The mean force potential for a grafted d-WLC hitting a hard wall normal to the grafting orientation}
\label{sec:WLC}

The very basic ingredient to introduce flexibility effects in the present modelling is the discrete-Wormlike model for passive grafted filaments. As justified in section \ref{sec:dynmodel} and appendix \ref{sec:char_time}, the action of individual grafted filaments pressing on a second obstacle can be treated adiabatically.
In ref. \cite{Frey.06} a semiflexible WLC of contour length $L_c$ tethered normally to a rigid surface at $x=0$ and impinging onto another rigid surface parallel to the first one at $x=L$ was investigated in the weak bending regime ($L\lesssim L_c$).  The average force exerted by the filament can be expressed as 
\begin{equation}
\bar{f}(L, L_c,\ell_p)=f_{b}(L_c,\ell_p)\tilde{f}(\eta)
\label{eq:frey}
\end{equation}
where the prefactor $f_{b}(L_c,\ell_p)$, the Euler buckling force, at fixed bending modulus $k_BT \ell_p$, goes like $L_c^{-2}$
\begin{equation}
f_{b}(L_c,\ell_p)=\frac{\pi^2}4\frac{k_BT\ell_p}{L_c^2}
\label{}
\end{equation}
and $\tilde{f}(\eta)$ is a universal function of the reduced compression $\eta=\left[(L_c-L)\ell_p/L_c^2\right]\geq 0$ ($L\leq L_c$). 
In ref. \cite{Frey.06} the following expression of $\tilde{f}(\eta)$ has been obtained for a continuous WLC in the small compression regime and tested against Monte Carlo simulations of a d-WLC model 
\begin{eqnarray}
\tilde{f}(\eta)&=&-\frac{4}{\pi^2} \frac{\partial \ln Z(\eta)}{\partial \eta} \label{eq:ftilde}\\
Z(\eta)&=&2\sum_{k=1}^{\infty} (-1)^{k+1}\frac{e^{-\lambda_k^2 \eta}}{\lambda_k},  \quad \lambda_k=(2k-1)\frac{\pi}{2}\label{eq:zetaeta}
\end{eqnarray}
where as usual $L_c=(j-1)d$ with $d$ the bond length and $j$ the number of articulation points in the chain. 
It was shown that $\tilde{f}(\eta)$ rapidly grows from $\tilde{f}(0)=0$ to the plateau value of 1 reached at $\eta\simeq 0.25$ and remains at this unitary value in a wide range of $\eta$ values. The behavior of continuous and discrete WLC models are in very good agreement provided $L\gg d$ and $L_c\gg d$ in the d-WLC.

Note that $\eta$ is proportional to $\ell_p$ hence the more rigid the filament, the larger $\eta$ at given $L_{c}-L$. In the case of actin ($\ell_p=5730d$), $\eta \simeq 0.25$ (or $\gamma/L_c \simeq 5\times 10^{-5}~d^{-1}$) corresponds to a compression $L_c-L\simeq 0.1 d$ for typical values of $L_c\simeq 50$. 
Hence the force exerted by a compressed actin filament is almost always equal to the $L-$independent Euler buckling force. 
Moreover for any polymerizing filament whose tip in its straight configuration is at a distance less than 0.9d from the obstacle before the chemical step, the force will jump from 0 to $f_b$ as the result to the addition of one bond. 

In reference \cite{PCR.15} the compressional force exerted by a  d-WLC filament upon a rigid wall has been studied by Monte Carlo simulations for values of $(L_c-L)$ well beyond the weak bending limit. 
The dimensionless force $\tilde{f}(\eta)$ was observed to deviate from its plateau value of 1 at a $\eta$ value inversely proportional to the chain contour length $L_c$. 
For increasing $\eta$ the force rapidly increases and it is expected to become independent of $L_c$ because the chain is strongly bent against the obstacle and only the fraction of the filament of length $\sim \pi L/2 \ll L_c$ closer to its tethering point remains under compression, while the remaining part is free to lay parallel to the obstacle wall (or better to undergo equilibrium fluctuations without crossing the obstacle wall). 

Introducing another reduced compression $\widetilde{\gamma}=(L_c-L)/L_c=\eta L_c/\ell_p$ plotting the large compression data for chains of different $L_c$ (but same $\ell_p=5370d$) as a function of $\widetilde{\gamma}$, they collapse on the same curve as shown in figure \ref{fig:Fig1SM}. 
This large compression behavior departs from the unitary plateau value at $\widetilde{\gamma}\simeq 0.1$. In the figure we also report the following fitting function which reproduces reasonably well this unique behavior
\begin{equation}
\tilde{\tilde{f}}(\widetilde{\gamma})=\frac{(a+b\widetilde{\gamma}^2)}{(1-\widetilde{\gamma})^2}+1 \hskip 1cm \widetilde{\gamma}\geq 0.1
\label{fit}
\end{equation}
with $a=0.044(5)$, $b=0.28(1)$. These values of the parameters have been obtained by fitting simultaneously the three sets of data in the figure.

The full expression of the adiabatic force exerted by a d-WLC of contour length $L_c$ upon a rigid wall at distance $L$ from its seed is given by:
\begin{equation}
\bar{f}(L,L_c,\ell_p)=f_b(L_c,\ell_p)\times
\begin{dcases}
\tilde{f}\left(\widetilde{\gamma} \frac{\ell_p}{L_c}\right) \qquad &\widetilde{\gamma} < 0.1\\
\frac{(a+b \widetilde{\gamma}^2)}{(1-\widetilde{\gamma})^2}+1 \qquad&\widetilde{\gamma}\geq 0.1
\end{dcases}
\label{eq:force_app}
\end{equation}

The large contour length expression -- second line of eq. (\ref{eq:force_app}) -- leads, in the infinite $L_c$ limit ($L/L_c\to0$, $\widetilde{\gamma}\to 1$), to the following asymptotic expression:
\begin{equation}
\lim_{L_{c}\to\infty}\bar{f}(L,L_c,\ell_p)=\frac{\pi^2}4k_BT\ell_p\frac{a+b}{L^2}
\end{equation}
For active filaments in supercritical conditions, this regime is attained rapidly after the filament escapes laterally. Indeed the propensity to polymerize at the tip of the filament and the loss of contact between the filament tip and the obstacle 
will rapidly result in an unlimited contour length growth to reach the asymptotic regime. 
In this regime the force exerted by the filament is related essentially to the elastic response due to pure bending and not to chemical activity at the tip of the filament. In an optical trap apparatus of strength $\kappa_T$, the mechanical equilibrium condition ($\bar{f}=\kappa_T L$) in this asymptotic regime imposes 
\begin{equation}
L_{esc}=\left(\frac{\pi^2}{4}k_BT\ell_p\frac{a+b}{\kappa_T}\right)^{1/3}
\label{eq:Lesc}
\end{equation}
For a bundle of $N_f$ escaped filaments this value must be multiplied by $N_f^{1/3}$.

$L_{esc}$ can be compared to $L_s=k_BT \ln(\hat{\rho}_1)/(d\kappa_T)$ a rather good approximation for the average width of the trap in the non escaping regime \cite{PPCR.16}. The condition $L_{esc}=L_s$ for a single filament sets a threshold $\bar{\kappa}_T$ for the trap strength to switch from one regime to the other:
\begin{equation}
\bar{\kappa}_T^2= \frac{4}{\pi^2} \frac{\left[\ln(\hat\rho_1)\right]^3}{(a+b)} \frac{(k_BT)^2}{\ell_p d^3}
\label{eq:kappa_esc}
\end{equation}
For $\kappa_T > \bar{\kappa}_T$ the filament will be non-escaping while for $\kappa_T < \bar{\kappa}_T$ the filament has high probability to escape and to end up in the asymptotic regime. Using the values of the parameters $a$ and $b$ given above, the actin persistence length and $\hat{\rho}_1=2.5$ with $k_BT=1$ and $d=1$ we get: $\bar{\kappa}_T=0.01337$. This value is indeed compatible with the value used in ref. \cite{PPCR.16} in the non-escaping regime. For a bundle of $N_f$ filaments the threshold value is multiplied by $N_f$. For $N_f=32$ we then obtain $\bar{\kappa}_T=0.4284$ compatible with the results shown in figure 4 of the main paper.

Eq. (\ref{eq:force}) fully characterize the force-compression law of the d-WLC model. Using this relation we can study the behavior of our "apparatus" (bundle+optical trap) for any number of filaments $N_f$ and any trap strength $\kappa_T$, in particular we can mimic the conditions of the experiments reported in ref \cite{Footer.07}. 

\end{appendix}
\bibliography{bibliography}
\end{document}